\documentclass[sn-basic]{sn-jnl}

\usepackage{graphicx}%
\usepackage{multirow}%
\usepackage{amsmath,amssymb,amsfonts}%
\usepackage{amsthm}%
\usepackage{mathrsfs}%
\usepackage[title]{appendix}%
\usepackage{xcolor}%
\usepackage{textcomp}%
\usepackage{manyfoot}%
\usepackage{booktabs}%
\usepackage{algorithm}%
\usepackage{algorithmicx}%
\usepackage{algpseudocode}%
\usepackage{listings}%

\theoremstyle{thmstyleone}%
\theoremstyle{thmstyletwo}%

\theoremstyle{thmstylethree}%

\newcommand{\actaa}{Acta Astron.}

\newcommand{\apj}{Astrophys. J.}
\newcommand{\apjl}{Astrophys. J. Lett.}
\newcommand{\apjs}{Astrophys. J. Suppl. Ser.}

\newcommand{\apss}{Astrophys. Space Sci.}
\newcommand{\aap}{Astron. Astrophys.}

\newcommand{\jcap}{J. Cosmol. Astropart. Phys.}

\newcommand{\mnras}{Mon. Not. R. Astron. Soc.}

\newcommand{\pasj}{Publ. Astron. Soc. Jpn}

\begin{document}
		
	\title[Relativistic hot accretion flow]{Properties of relativistic hot accretion flow around rotating black hole with radially varying viscosity}

	\author{\fnm{Monu} \sur{Singh}}\email{monu18@iitg.ac.in}
	\equalcont{These authors contributed equally to this work.}
	
	\author*{\fnm{Santabrata} \sur{Das}}\email{sbdas@iitg.ac.in}
	\equalcont{These authors contributed equally to this work.}
	
	\affil{\orgdiv{Department of Physics}, \orgname{Indian Institute of Technology Guwahati}, \orgaddress{\city{Guwahati}, \postcode{781039}, \state{Assam}, \country{India}}}
		
	\date{\today}



	
	\abstract{We examine the effect of variable viscosity parameter ($\alpha$) in relativistic, low angular momentum advective accretion flow around rotating black holes. Following the recent simulation studies of magnetohydrodynamic disk that reveal the radial variation of $\alpha(r)$, we theoretically investigate the properties of the global transonic accretion flow considering a one-dimensional power law prescription of viscosity parameter as $\alpha(r) \propto r^{\theta}$, where the viscosity exponent $\theta$ is a constant. In doing so, we adopt the relativistic equation of state and solve the fluid equations that govern the flow motion inside the disk. We find that depending on the flow parameters, accretion flow experiences centrifugally supported shock transition and such shocked accretion solutions continue to exist for wide ranges of the flow energy, angular momentum, accretion rate and viscosity exponent, respectively. Due to shock compression, the hot and dense post-shock flow (hereafter PSC) can produce the high energy radiations after reprocessing the soft photons from the pre-shock flow via inverse Comptonization. Since PSC is usually described using shock radius ($r_s$), compression ratio ($R$) and shock strength ($S$), we study the role of $\theta$ in deciding $r_s$, $R$ and $S$, respectively. Moreover, we obtain the parameter space for shock and find that possibility of shock formation diminishes as $\theta$ is increased.	Finally, we compute the limiting value of $\theta$ ($i.e., \theta^{\rm max}$) that admits shock and find that flow can sustain more viscosity when it accretes onto rapidly rotating ($a_{\rm k} \rightarrow 1$) black hole in comparison to weakly rotating ($a_{\rm k} \rightarrow 0$) black hole.}

\maketitle


\section{Introduction}

Accretion of matter onto compact object is considered to be the most efficient energy release process. However, in the context of accretion disk theory, the underlying mechanisms responsible to transport the angular momentum through the disk are not yet well understood and remain the most intriguing unresolved problem due to the disagreement of findings between the numerical simulations \cite[]{Balbus-Hawley1998} and observational results \cite[]{Smak1999}. In particular, \cite{King-etal2007} found an apparent discrepancy of a factor of $\sim 10$ between observational and theoretical estimates of viscosity parameter in accretion flow around black hole (BH).

In a seminal work, \cite{Shakura-Sunyaev1973} (hereafter SS73) introduced dimensionless viscosity parameter $\alpha$ defined as the ratio of the viscous stress to the pressure of the accretion flow. In the absence of detailed understanding of the viscous mechanism, SS73 considered $\alpha$ to be a global constant all throughout, typically in the range $0.001-0.1$ \cite[and references therein]{Das-etal2021}. Afterwords, considering the effective shear viscosity driven by the magneto-rotational instability, \cite{Hawley-Krolik2001,Hawley-Krolik2002} suggested that $\alpha$ may not be constant throughout the flow, instead it possibly varies both spatially and temporally in an accretion flow. Similar findings were also reported by numerous groups of researchers while examining the overall characteristics of $\alpha$ using magneto-hydrodynamical simulations \cite[]{Balbus-Hawley1991,Matsumoto-Tajima1995,Hawley-etal1995,Hawley-etal1996,Lyubarskii-1997,Nayakshin-1999,Steinacker-Papaloizou2002,Sano-etal2004,Fragile-etal2007,Penna-etal2010,Penna-etal2012,Porth-etal2019}. Very recently, \cite{Mitra-etal2022} computed the profile of the ratio of Maxwell stress to the gas pressure, and found that unlike standard SS73 viscosity parameter, it also varies with radial distance as well. Needless to mention that the measure of $\alpha$ from both local and global simulations is undoubtedly challenging as it depends on several factors, namely initial magnetic field geometry and strength, grid resolutions, etc \cite[]{Sorathia-etal2012}. Accordingly, it appears that the range of $\alpha$ values is not well constrained and hence, it remains inconclusive. 

Indeed, it is the viscous stress that generally varies inside the disk, and hence, $\alpha$ is often considered as radially varying. Adopting these ideas, we investigate the properties of relativistic viscous advective accretion flow around the rotating BHs. During accretion, matter starts accreting with subsonic speed from a large distance and plunges into {\rm BH} supersonically to meet the horizon condition. Because of this, inflowing matter experiences smooth transition from sub- to super-sonic domain at least once while accreting onto BH. However, flow can encounter such sonic transitions multiple times depending on the flow energy and angular momentum, and solution of this kind is specially encouraging because centrifugal wall may trigger shock transitions \cite[]{Fukue-1987,Chakrabarti-1989,Abramowicz-etal1990,Yang-Kafatos-1995,Das-etal2001a,Das-etal2001b,Chakrabarti-Das2004,Das-2007,Kumar-Chattopadhyay2014,Sarkar-Das2016,Das-Sarkar2018,Dihingia-etal2018,Dihingia-etal2019,Dihingia-etal2019a,Das-etal2021,Sen-etal2022,Patra-etal2022}. Such shock transitions are possible provided Rankine-Hugonoit conditions (RHCs) are favourable \cite[]{Landau-Lifshitz1959}. At shock, accreting matter jumps from supersonic to subsonic value and this causes the convergent post-shock matter hot, dense, and puffed-up, which is commonly referred as post-shock corona (PSC). After the shock, accreting matter starts moving towards the horizon and gradually gains radial velocity. This process continues and finally, matter crosses the event horizon ($r_g$) with supersonic speed after passing through a critical point located close to $r_g$. In reality, PSC comprises swarm of the hot electrons. These electrons interact with the soft photons from pre-shock matter via inverse-Comptonization process and produce hard X-ray radiations \cite[]{Chakrabarti-Titarchuk1995,Mandal-Chakrabarti2005}. When RHCs do not satisfied, however flow possesses more than one critical point, PSC is expected to exhibit time varying modulation that usually may give rise to the quasi-periodic variations of emitted photons commonly observed from Galactic BH sources \cite[and references therein]{Chakrabarti-Manickam2000,Nandi-etal2001,Nandi-etal2012,Majumder-etal2022,Das-etal2021}.

Being motivated with this, we examine the structure of the steady, viscous, advective flow that accretes on to a rotating BH. We adopt the viscosity parameter that is radially varying as $\alpha(r) = \alpha_{0} (r/r_{g})^{\theta}$, where $r$ is the radial distance, $r_{g}$ is gravitational radius, $\alpha_{0}$ is the proportionality constant and $\theta$ is the viscosity exponent. We consider relativistic equation of state (REoS) that satisfactorily accounts the thermal properties of the low angular momentum accreting matter \cite[]{Chattopadhyay-Ryu2009,Dihingia-etal2019}. Further, we use a recently developed pseudo potential to mimic the BH gravity \cite[]{Dihingia-etal2018} for spin values ranging from weakly rotating ($a_{\rm k}\rightarrow 0$) to rapidly rotating ($a_{\rm k}\rightarrow 1$) limits. Considering all these, we calculate the global transonic accretion solutions (GTAS) by solving the fluid equations using the accretion model parameters. Moreover, we identify the requisite GTAS that admit standing shock transitions, and render the dependencies of the dynamical as well as thermodynamical flow variables on the model parameters. In addition, we investigate the various shock properties, such as shock radius($r_s$), compression ratio($R$), and shock strength($S$), respectively, and study how $r_s$, $R$ and $S$ depend on the viscosity ($\alpha$) and accretion rate ($\dot m$). We also determine the range of flow energy($\cal E$) and angular momentum($\lambda$) that render shock-induced GTAS, and ascertain the domain of shock parameter space in $\lambda-\cal E$ plane. We find that such a parameter space is altered as the viscosity exponent ($\theta$) is varied. Since $\theta$ plays pivotal role in deciding the accretion disc structure, it is important to explore the limiting value of the viscosity exponent ($\theta^{\rm max}$). Accordingly, we put effort to estimate $\theta^{\rm max}$ and find that it fervently depends on both BH spin ($a_{\rm k}$) and $\alpha_0$.

This paper is organized in the following manner. In \S2, we describe the model considerations and basic equations. We obtain GTAS both in absence and presence of shock and discuss the shock properties in \S3. In \S 4, we discuss how the shock parameter space alters with the change of viscosity and calculate the maximum limit of viscosity exponent for shock. Finally, in \S 5, we present conclusions.

\section{Basic considerations and model equations}

We begin with an axisymmetric, steady-state, height-averaged viscous advective accretion disk around a rotating BH in presence of synchrotron cooling. We also consider that such a disk remains confined around the disk equatorial plane. We approximate the effect of gravity by adopting an effective pseudo-potential \cite[]{Dihingia-etal2018} that successfully delineates the spacetime wrapping due to rotating BH. The accretion process inside the disk is driven by the viscous stress ($W_{r\phi}$) and we consider $W_{r\phi} = -\alpha \Pi$, where $\Pi$ denotes the vertically integrated pressure including ram pressure, and $\alpha$, the viscosity parameter, is the dimensionless quantity that is assumed to vary with radial coordinate. Needless to mention that $\alpha$ absorbs all the detailed microphysics of the viscous processes. With these considerations, we express all the governing equations using $M_{\rm BH}=G=c=1$, where $M_{\rm BH}$, $G$ and $c$ denote BH mass, gravitational constant and light speed, respectively. With this, we write the length in unit of $r_g=GM_{\rm BH}/c^2$, and accordingly, time and angular momentum are written in unit of $r_g/c$ and $r_g c$.

\subsection{Governing Equations}

The basic fluid equations that describe the motion of the accreting matter inside the disc around a rotating BH are as follows:

\noindent (a) Conservation equation for radial momentum:
\begin{equation}
	\upsilon \frac{d\upsilon}{dr}+\frac{1}{h \rho}\frac{dP}{dr}+\frac{d \Phi_{\textrm{e}}^{\textrm{eff}}}{dr} = 0,
	\label{Eq:1}
\end{equation}
where $\upsilon$, $P$, $\rho$ and $h$, denote the flow velocity, gas pressure, mass density and specific enthalphy, respectively. In addition, $\Phi_{\rm e}^{\rm eff}$ refers the effective potential of a rotating BH that mimics the spacetime geometry at the disc equatorial plane and is given by \cite[]{Dihingia-etal2018},
$$
\Phi_{\textrm{e}} ^{\textrm{eff}} = \frac{1}{2}\ln\left[\frac{r \Delta}{a_{\textrm{k}}^2 (r+2) - 4 a_{\textrm{k}} \lambda + r^3 -\lambda^2(r-2)}\right],
\eqno(1a)
$$
where $\lambda$ is the specific angular momentum of the accreting matter, $a_{\rm k}$ denotes the BH spin, and $\Delta = r^2 - 2 r + a_{\rm k}^2$.
	
\noindent (b) Mass conservation equation:
\begin{equation}
	 \dot{M}= 2\pi \upsilon \Sigma \sqrt{\Delta}\thickspace,
	 \label{Eq:2}
\end{equation}
where $\dot{M}$ is the mass accretion rate. In this work, we do not consider the ejection of matter in the form of outflow/jets and hence, $\dot{M}$ is treated as global constant in absence of any mass loss from the disk. Moreover, for convenience, we express the accretion rate in unit of Eddington accretion rate as ${\dot m} = {\dot M}/{\dot M}_{\rm Edd}$, where ${\dot M}_{\rm Edd} = 1.44 \times 10^{17}\left( \frac{M_{\rm BH}}{M_\odot}\right)$. Further, $\Sigma$ is the vertically integrated surface mass density of the accreting matter \cite[]{Matsumoto-etal1984}, and is written as $\Sigma = 2 \rho H$, where $H$ {refers} the local disc half thickness expressed as \cite[]{Riffert-Herold1995,Peitz-Appl1997},
$$
H^2 = \frac{P r^{3}}{\rho {\cal F}}, \thickspace {\cal F} = \gamma_{\phi}^{2} \frac{(r^2 + a_{\textrm{k}}^{2})^{2} + 2 \Delta a_{\textrm{k}}^{2}}{(r^2 + a_{\textrm{k}}^{2})^{2} - 2 \Delta a_{\textrm{k}}^{2}}, \thickspace  \gamma_{\phi}^{2} = \frac{1}{(1-\lambda \Omega)} \thickspace.
$$
Here, $\Omega ~ \left[ = (2 a_{\rm k}+\lambda(r-2))/(a_{\rm k}^{2}(r + 2) - 2 a_{\rm k}\lambda + r^3)\right]$ denotes the angular velocity of the accreting matter.

\noindent (c) Conservation equation for azimuthal momentum:
\begin{equation}
	\upsilon \frac{d\lambda}{dr} + \frac{1}{\Sigma\space r}\frac{d}{dr}(r^2 W_{r\phi}) = 0,
	\label{Eq:3}
\end{equation}
where, we consider the $r\phi$ component of the viscous stress as $ W_{r\phi} = - \alpha\Pi = - \alpha (W + \Sigma \upsilon^2)$ \cite[][and references therein]{Chakrabarti-Molteni1995,Chakrabarti-Das2004}. In equation \ref{Eq:3}, $W$ denotes the vertically integrated pressure and $\Sigma$ represents the vertically integrated mass density. In this work, we consider radially varying viscosity parameter resembling power law distribution as 
$$
\alpha = A \left( \frac{r}{r_g}\right)^{\theta}\thickspace  = \alpha_{0} r^{\theta},
\eqno(3a)
$$
where $A$, $\theta$ and $\alpha_{0}$ are regarded as constants all throughout the flow. Similar findings on the radial variation of viscosity parameter are also recently reported by the group of workers \cite[]{Penna-etal2013,Zhu-Stone2018}. Note that when $\theta \rightarrow 0$, we obtain globally constant viscosity parameter $\alpha = \alpha_{0}$, as in the case of `$\alpha$ model' prescription \cite[][]{Shakura-Sunyaev1973}.

\noindent (d) Equation for energy balance:
\begin{equation}
	\Sigma \upsilon T \frac{ds}{dr} = \frac{\upsilon H}{\Gamma -1}\left( \frac{dP}{dr} - \frac{\Gamma P}{\rho}\frac{d\rho}{dr}\right)= Q^{-}-Q^{+}.
	\label{Eq:4}
\end{equation}
In equation (\ref{Eq:4}), $T$ is the flow temperature, $s$ is the specific entropy and and $\Gamma$ is the adiabatic index. Moreover, during accretion, the heat gain and lost by the flow are denoted by $Q^{+}$ and $Q^{-}$, respectively. Following \cite{Chakrabarti1996,Aktar-etal2017}, we adopt the mixed shear stress prescription to compute the viscous heating of the flow and is given by,
\begin{equation}
	Q^{+}=-\alpha \rho H r \left(\frac{P}{\rho}+\upsilon^2\right)\frac{d\Omega}{dr}.
	\label{Eq:5}
\end{equation} 

In general, the bremsstrahlung cooling process is regarded as the inefficient cooling process \cite[]{Chattopadhyay-Chakrabarti2000}. Hence, in this work, we consider energy loss due to synchrotron cooling only. Accordingly, the synchrotron emissivity of the convergent accretion flow is obtained as \cite[]{Shapiro-Teukolsky1983},
\begin{equation}
	Q^{-} = Q^{\rm syn} =\frac{16}{3}\frac{e^2}{c}\left(\frac{e B}{m_{e} c}\right)^2\left(\frac{k_{B} T}{m_{e} c^2}\right)^2 n_{e} \thickspace\thickspace \thickspace \textrm{erg cm$^{-3}$ s$^{-1}$},
	\label{Eq:6}
\end{equation}
where $e$, $m_{e}$, and $n_{e}$ are the charge, mass, and number density of the electrons respectively, $k_{B}$ is the Boltzmann constant, $B$ is the magnetic fields. In the astrophysical context, the present of magnetic fields is ubiquitous inside the disc and hence, the ionized flow should emit synchrotron photons causing the accreting flow to cool down significantly. Indeed, the characteristics of structured magnetic fields inside the disc still remains unclear, and hence, we rely on the random or stochastic magnetic field. For the purpose of simplicity, we use equipartition to estimate magnetic field and obtained as, $B = \sqrt{8 \pi \beta P}$, where $\beta$ is a dimensionless constant. Evidently, $\beta \lesssim 1$ confirms that the magnetic fields remain confined within accretion disc \cite[]{Mandal-Chakrabarti2005}. For the purpose of representation, in this work, we choose $\beta = 0.1$.

In order to close the governing equations (\ref{Eq:1} - \ref{Eq:4}), one requires to consider an equation of state (EoS) that relates $P$, $\rho$ and internal energy ($\epsilon$) of the flow. Hence, we consider an EoS for relativistic flow which is given by \cite[]{Chattopadhyay-Ryu2009},
\begin{equation}
	\epsilon = \frac{\rho f}{\left(1+\frac{m_p}{m_e}\right)},
	\label{Eq:7}
\end{equation}
with
$$
f = \left[ 1+\Theta \left( \frac{9\Theta + 3}{3 \Theta +2} \right) \right] + \left[ \frac{m_p}{m_e} + \Theta \left( \frac{9 \Theta m_e + 3 m_p}{3 \Theta m_e + 2 m_p} \right) \right],
$$
where $\Theta ~(=k_B T/m_e c^2)$ is the dimensionless temperature of the flow. Utilizing the relativistic EoS, we express polytropic index as $N = \frac{1}{2}\frac{df}{d\Theta}$, adiabatic index as $\Gamma = 1+ 1/N$ and sound speed as $C_{s}^2 = \frac{\Gamma P}{e + P} = \frac{2\Gamma \, \Theta}{f + 2\Theta}$, respectively \cite[and references therein]{Mitra-etal2022}. Further, following \cite{Chattopadhyay-Kumar2016} and with the help of equation (\ref{Eq:2}), we compute the entropy accretion rate as $ \dot{\mathcal{M}} = \upsilon H \sqrt{\Delta} \left[ \Theta^2\left(2+3\Theta\right)\left(3\Theta + 2m_p/m_{e}\right) \right]^{3/4} \exp(k_{1}) $, where $k_{1} = 0.5 \times \left[f/\Theta - (1+ m_{p}/m_{e})/\Theta\right]$.

Using equations (\ref{Eq:1}-\ref{Eq:7}), we get the radial velocity gradient in the form of wind equation as,
\begin{equation}
\frac{d\upsilon}{dr} = \frac{{\cal N}(r, \upsilon, \Theta, \lambda, \alpha )}{{\cal D}(r, \upsilon, \Theta, \lambda, \alpha )},
\label{Eq:8}
\end{equation}
where both ${\cal N}$ and ${\cal D}$ depend on $r, \upsilon, \Theta, \lambda, \alpha$, and their explicit mathematical expressions are given in Appendix A. Using equation (\ref{Eq:8}), we obtain the radial derivatives of angular momentum ($\lambda$) and dimensionless temperature ($\Theta$) as,
\begin{equation}
\frac{d\lambda}{d r} = \lambda_{1} + \lambda_{2}\frac{d\upsilon}{d r},
\label{Eq:9}
\end{equation}
and
\begin{equation}
\frac{d\Theta}{d r} = \Theta_{1} + \Theta_{2}\frac{d\upsilon}{d r},
\label{Eq:10}
\end{equation}
where the mathematical form of the coefficients, such as $\lambda_{1}$, $\lambda_{2}$, $\Theta_{1}$, and $\Theta_{2}$ are described in Appendix A.

Indeed, during accretion, subsonic flow commences accreting towards the BH from a far away distance $r_{\rm edge}$ (hereafter disc outer edge) and crosses BH horizon ($r_h$) supersonically. Therefore, flow ought to pass through a critical point ($r_c$) where it smoothly transits from subsonic to supersonic domain. Note that for $r > r_h$, flow may possess multiple critical points depending on the flow parameters. Following \cite{Chakrabarti-Das2004}, we carry out the critical point analysis, where $\left(\frac{d\upsilon}{dr}\right)_{r_c}=\frac{0}{0}$ as one gets ${\cal N} = {\cal D}= 0$ at the critical point. Because of this, we make use of the l\'{}Hospital rule while computing $\left(\frac{d\upsilon}{dr}\right)$ at $r_c$. For physically acceptable solutions around BH, we consider saddle type critical points only where $\left(\frac{d\upsilon}{dr}\right)$ yields two distinct real values at the critical point \cite[and references therein]{Das-2007}. When $r_c$ forms near $r_h$, we refer it as inner critical point ($r_{\rm in}$), otherwise it is termed as outer critical point ($r_{\rm out}$) \cite[]{Chakrabarti-Das2004}.

\section{Global transonic accretion solutions (GTAS)}

\begin{figure}
	\begin{center}
		\includegraphics[width=0.9\textwidth]{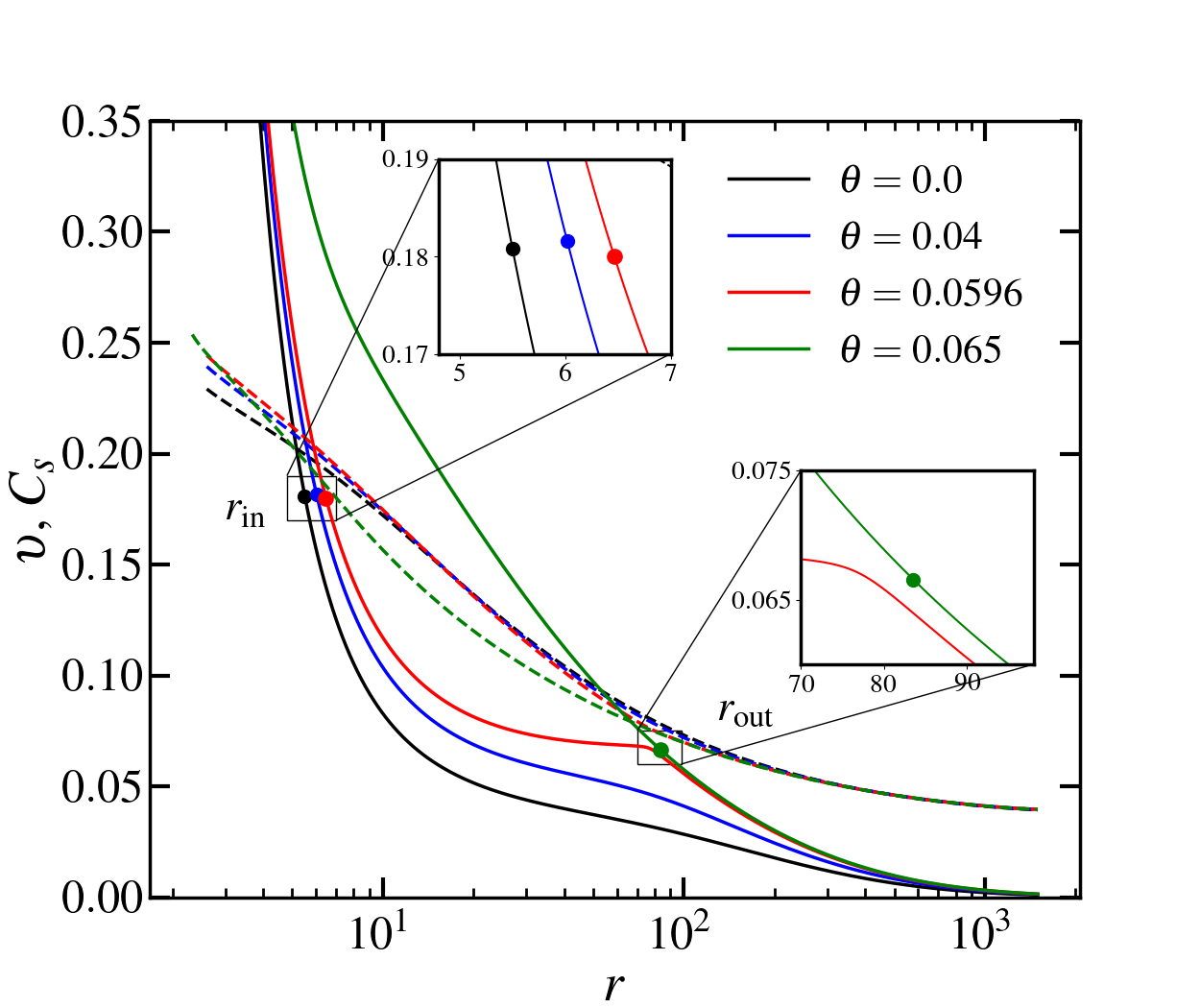}
		\caption{Plot of flow velocity ($\upsilon$) and sound speed ($C_{s}$) with radial distance ($r$) for different values of $\theta$. Dashed curves represent the sound speed and solid curves denote flow velocity. See the text for details.}
		\label{fig:1}
	\end{center}	
\end{figure}

\begin{table*}
	\caption{Power law exponent ($\theta$), critical point location ($r_{\rm c}$), critical point angular momentum ($\lambda_{\rm c}$), critical point velocity ($\upsilon_{\rm c}$), critical point temperature ($\Theta_{\rm c}$), disk outer edge ($r_{\rm edge}$), angular momentum at $r_{\rm edge}$ ($\lambda_{\rm edge}$), velocity at $r_{\rm edge}$ ($\upsilon_{\rm edge}$), temperature at $r_{\rm edge}$ ($\Theta_{\rm edge}$) for global transonic solutions presented in figure \ref{fig:1}. See the text for more details.
	}
	\begin{center}\label{table-1}
		\begin{tabular}{l c c c c c c c c l   }\hline\hline
			$\theta$ & $r_{\rm c}$ &  $\lambda_{\rm c}$   & $\upsilon_{\rm c}$ & $\Theta_{\rm c}$ & $r_{\rm edge}$ & $\lambda_{\rm edge}$ & $\upsilon_{\rm edge}$ & $\Theta_{\rm edge}$&  \\ 
			& $(r_g)$    & $(r_g c)$      & $(c)$     &$(m_e c^2 K)$ & $(r_g)$    & $(r_g c)$      & $(c)$     &$(m_e c^2 K)$ \\ \hline
			0 & 5.50 & 3.15 & 0.1808 & 27.9476 & 1500.0 & 21.10 & 0.00089 & 0.98115 \\
			
			0.04 & 6.019 & 3.03 & 0.1816 & 28.2146 & 1500.0 & 21.10 & 0.00118 & 0.98106 \\
			
			0.0596 & 6.464 & 2.96 & 0.1800 & 27.6714 & 1500.0 & 21.10 & 0.00137 & 0.98099 \\
			
			0.065 & 83.488 & 2.83 & 0.0665 & 3.4779 & 1500.0 & 21.10 & 0.00140 & 0.98098 \\
			\hline
		\end{tabular}
	\end{center}
	Note: Suffix `c' identifies quantities measured at inner (outer) critical point $r_{\rm in}$ ($r_{\rm out}$).
\end{table*}

The global transonic accretion solutions (GTAS) are obtained by solving the coupled differential equations (\ref{Eq:8} - \ref{Eq:10}) for a set of model parameters. Some of these parameters, namely $\alpha_0$, $\theta$, ${\dot m}$ and $a_{\rm k}$ remain constant all throughout, while the others, $i.e.$, critical point $r_c$ and angular momentum $\lambda_c$ at $r_c$ are treated as local parameters. Using the model parameters, we first integrate equations (\ref{Eq:8}-\ref{Eq:10}) starting from $r_c$ upto $r_h$ and again from $r_c$ to $r_{\rm edge} ~(\sim 1500$). Thereafter, a complete GTAS around BH is obtained by joining both parts of the solutions. Based on the choice of the model parameters, flow becomes transonic either at $r_{\rm in}$ or at $r_{\rm out}$ before crossing the BH horizon.

Figure \ref{fig:1} shows the typical sets of global transonic accretion solutions (GTAS) for flows injected from $r_{\rm edge}=1500$ with various $\theta$ values. Following the procedure mentioned in \cite{Das-2007}, we calculate the accretion solution containing the inner critical point $r_{\rm in}=5.50$, where we choose $\lambda_{\rm in}=3.15$, $\alpha_{0}=0.01$, $\theta = 0.0$, $a_{\rm k}=0.0$, and ${\dot m}=0.01$. This renders a global accretion solution as it successfully connects BH horizon $r_{h}$ with $r_{\rm edge}$. We note the flow variables at $r_{\rm edge}$ as $\lambda_{\rm edge}=21.10$, $\upsilon_{\rm edge}=8.9 \times 10^{-4}$, and $\Theta_{\rm edge}=0.98115$. In reality, we can get the same accretion solution once the flow equations are integrated towards BH horizon using these noted boundary values. Here, black solid curve denote the radial velocity $\upsilon(r)$, whereas black dashed curves represent the sound speed $C_s(r)$ of the flow for $\theta=0.0$. Next, we increase $\theta=0.04$ while keeping other flow variables unchanged at $r_{\rm edge}$ and calculate GTAS by suitably tuning $\upsilon_{\rm edge}=1.18 \times 10^{-3}$, and $\Theta_{\rm edge}=0.98106$. Here, we additionally require the boundary values of $\upsilon_{\rm edge}$ and $\Theta_{\rm edge}$ to integrate the fluid equations from $r_{\rm edge}$, as critical point remains unknown. The solution is depicted using blue color where we find that the inner critical point is shifted outwards as $r_{\rm in}=6.019$ for $\theta = 0.04$. Similarly, for $\theta=0.0596$, flow solution (red) continues to maintain similar character as in the case of $\theta=0.0$ and $0.04$, having inner critical points at $r_{\rm in}=6.464$. Solutions of this kind that are passing through $r_{\rm in}$ are similar to ADAF-type accretion solutions \cite[]{Narayan-Yi1994}. When $\theta$ is increased further, the nature of the flow solution (in green) is changed and it becomes transonic at outer critical point $r_{\rm out}=83.488$ rather than $r_{\rm in}$. When $\theta$ is increased further, flow solution (in green) changes its character and becomes transonic at the outer critical point $r_{\rm out}=83.488$ instead of the inner critical point. Usually, the solutions containing $r_{\rm out}$ are of Bondi type \cite[]{Bondi1952}. For the purpose of clarity, regions around the critical points ($r_{\rm in}$ and $r_{\rm out}$) are zoomed which are shown using filled circles at the insets. We tabulate the flow variables at $r_{\rm edge}$, $r_{\rm out}$, and $r_{\rm in}$ in Table \ref{table-1}. Overall, we observe that the role of $\theta$ is pivotal in deciding characteristic of GTAS around BHs.

\begin{figure}
	\begin{center}
		\includegraphics[width=0.7\textwidth]{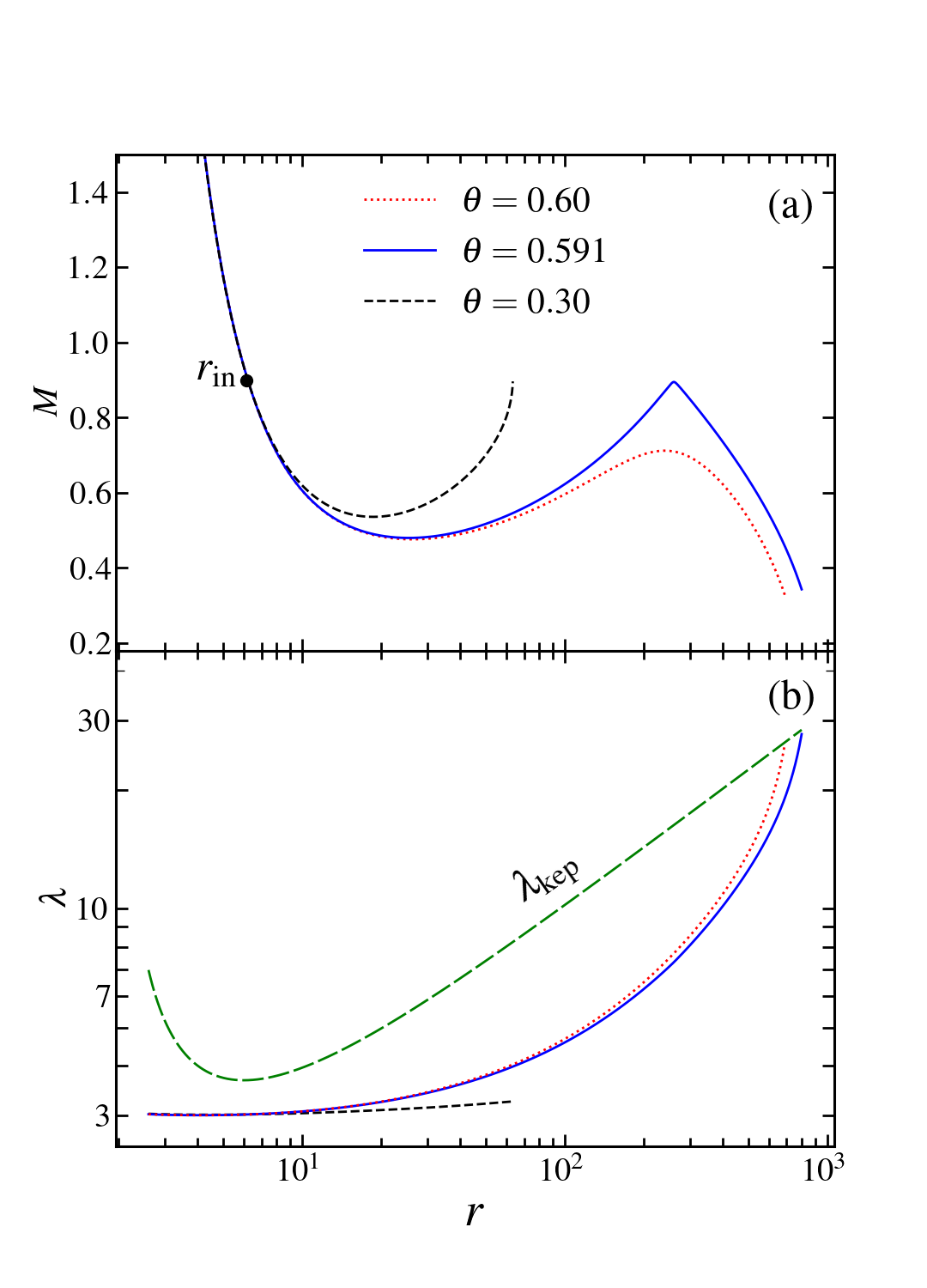}
		\caption{Variation of (a) Mach number $M~(=\upsilon/C_s)$ and (b) angular momentum ($\lambda$) with the radial distance $r$ for different $\theta$. Here, we choose $r_{\rm in} = 6.17$, $\lambda_{\rm in} = 3.01$, $\alpha_{0} = 0.01$, and ${\dot m}=0.01$. Dashed, solid and dotted curves denote results for $\theta =0.30,0.591$, and $0.60$, respectively. See the text for details.
		}
		\label{fig:2}
	\end{center}
\end{figure}

Next, we present the accretion flow solutions in figure \ref{fig:2}a, where the radial variation of Mach number ($M=\upsilon/C_s$) is demonstrated. Here, all the solutions become transonic at $r_{\rm in}=6.17$ with $\lambda_{\rm in}=3.01$, $\alpha_0=0.01$, ${\dot m} = 0.01$, $a_{\rm k}=0.0$, respectively. For $\theta = 0.6$, we obtain a GTAS that smoothly connects the BH horizon with $r_{\rm edge}$ where flow angular momentum matches with its Keplerian value as shown in dotted curve. We gradually decrease $\theta$ and find that beyond the limiting value as $\theta = 0.591$, accretion flow becomes closed as shown using solid curve. The result plotted using dashed curve corresponds to $\theta = 0.3$. The closed accretion solutions passing through $r_{\rm in}$ are noteworthy as they can join with another solution passing through $r_{\rm out}$ via centrifugally supported shocks. Indeed, the existence of shocks in advective accretion flows has intense implication because the solution of this kind satisfactorily explains the temporal and spectral properties of BH sources \cite[]{Molteni-etal1994,Molteni-etal1996,Chakrabarti-Titarchuk1995,Chakrabarti1996,Lu-etal1999,Chakrabarti-Manickam2000,Das-etal-2009,Nagakura-Yamada2009,Nandi-etal2012,Iyer-etal2015,Okuda-Das2015,Sukova-Janiuk2015,Das-etal2021}. Accordingly, in the subsequent sections, we investigate the shock-induced GTAS around BHs. In figure \ref{fig:2}b, we present the variation of angular momentum for the solutions presented in figure \ref{fig:2}a, where big-dashed curve denote the Keplerian angular momentum profile.

\begin{figure}
	\begin{center}
		\includegraphics[width=0.7\textwidth]{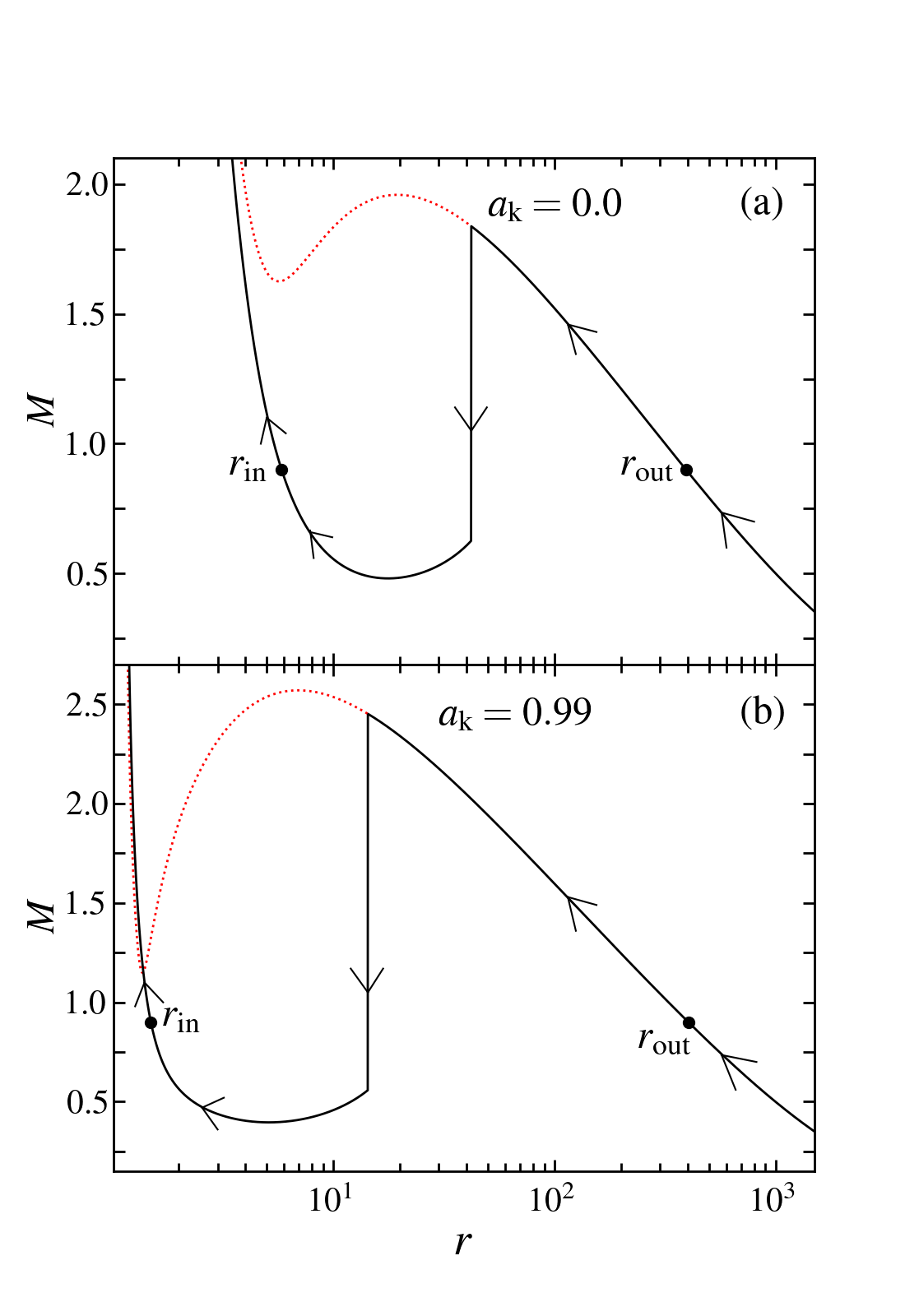}
		\caption{Examples of shock-induced GTAS around BH, where variation of $M$ with $r$ are depicted. Vertical down-arrow shows the shock transition radius, where Rankine-Hogoniot conditions \citep{Landau-Lifshitz1959} for standing shock are satisfied. Filled circles denote critical points and arrows show the direction of the flow motion towards BH. Dotted curve represents shock free solution. (a) Upper panel is for $a_{\rm k} = 0.0$, and (b) lower panel is for $a_{\rm k} = 0.99$. See the text for details.
		}
		\label{fig:3}
	\end{center}
\end{figure}

In figure \ref{fig:3}a, we depict an example of shock-induced global accretion solution, which passes thorough both $r_{\rm out}=395.32$ and $r_{\rm in}=5.808$ while accreting onto a stationary BH ($a_{\rm k}=0.0$). Here, matter is injected sub-sonically from $r_{\rm edge}=1500$ with $\lambda_{\rm edge}=4.68$, $\upsilon_{\rm edge}=8.35\times 10^{-3}$, $\Theta_{\rm edge}=0.336$, ${\dot m}=0.01$, $\alpha_0 = 0.01$, and $\theta=0.1$. As subsonic matter proceeds towards the BH, it becomes supersonic at $r_{\rm out}=395.32$ and proceeds further towards the BH. Indeed, accreting matter can seamlessly cross the BH horizon after passing $r_{\rm out}$ as shown using dotted curve. Interestingly, supersonic accreting matter sees an alternative possibility of discontinuous shock transition of the flow variables in the subsonic branch as the Rankine-Hogoniot conditions (RHCs) \cite[]{Landau-Lifshitz1959} for standing shock are satisfied at shock radius ($r_s$). We determine the standing shock location $r_s=42.07$ for a vertically integrated flow by employing shock conditions (RHCs) which are (a) continuity of energy flux: $[{\cal E}]=0$, (b) continuity of mass flux: $[\dot{M}]=0$, and (c) continuity of momentum flux $[W + \Sigma \upsilon^2] =0$ across shock. Here, we express the local energy of the flow as ${\cal E}=\upsilon^2/2+\log h + \Phi^{\rm eff}_{\rm e}$, and the quantities within the bracket (\verb|[ ]|) denote their differences across shock transition location. We show the shock transition using vertical arrow. Immediately after shock transition, the radial velocity of the matter decelerates, however, it  progressively increases as the matter proceeds towards the horizon. Eventually, matter crosses the BH horizon at supersonic speed after passing through the inner critical point at $r_{\rm in}=5.808$. In the figure, arrows show how the matter moves towards BH. Note that the post-shock branch of shocked solution is similar in nature to the solution for $\theta = 0.3$ in figure \ref{fig:2}. Further, we calculate the shock-induced GTAS around a rotating BH of $a_{\rm k}=0.99$ for flows injected with $\lambda_{\rm edge}=3.68$, $\upsilon_{\rm edge}=8.33 \times 10^{-3}$, $\Theta_{\rm edge}=0.342$, ${\dot m}=0.01$, $\alpha_0 = 0.01$, and $\theta=0.1$ from $r_{\rm edge}=1500$. The shock is formed at $r_s = 14.33$ in between $r_{\rm out}=405.46$ and $r_{\rm in}=1.492$. Here, we observe that for a chosen set of ($\alpha_0,\theta$), shock exists around rapidly rotating BH when $\lambda_{\rm edge}$ assumes relatively smaller value and vice versa. Indeed, this is expected because accreting matter crosses the BH horizon with angular momentum lower than the marginally stable angular momentum ($\lambda_{\rm ms}$) and $\lambda_{\rm ms}$ evidently decreases with the increase of $a_{\rm k}$ \citep{Das-Chakrabarti2008}. In Table \ref{table-2}, we present the flow variables at the critical points for shock-induced GTAS presented in Figure \ref{fig:3}.

\begin{table*}
	\caption{Black hole spin ($a_{\rm k}$), critical point location ($r_{\rm c}$), angular momentum ($\lambda_{\rm c}$), radial velocity ($\upsilon_c$), and temperature ($\Theta_{\rm c}$) measured at $r_{\rm c}$ for shocked accretion solution presented in figure \ref{fig:3}. Subscript `c' identifies quantities measured either at inner (`in') or outer (`out') critical point. See text for more details.}
	\begin{center}
		\label{table-2}
		\begin{tabular}{c c c  c  c  c  l   }\hline\hline
			$a_{\rm k}$ & Critical & $r_{\rm c}$      & $\lambda_{\rm c}$     & $\upsilon_{\rm c}$  & $\Theta_{\rm c}$&  \\ 
			& Point    & $(r_g)$    & $(r_g c)$      & $(c)$     &$(m_e c^2 K)$ \\ \hline
			0 & Inner & 5.808 & 3.091 & 0.1794 & 27.4538 \\
			
			& Outer & 395.32 & 3.470 & 0.0313 & 0.7580 \\ \hline
			
			0.99 & Inner & 1.492 & 2.058 & 0.2653 & 68.1666 \\
			
			& Outer & 405.46 & 2.445 & 0.0312 & 0.7510 \\
			\hline
		\end{tabular}
	\end{center}
\end{table*}

\begin{figure}
	\begin{center}
		\includegraphics[width=0.9\textwidth]{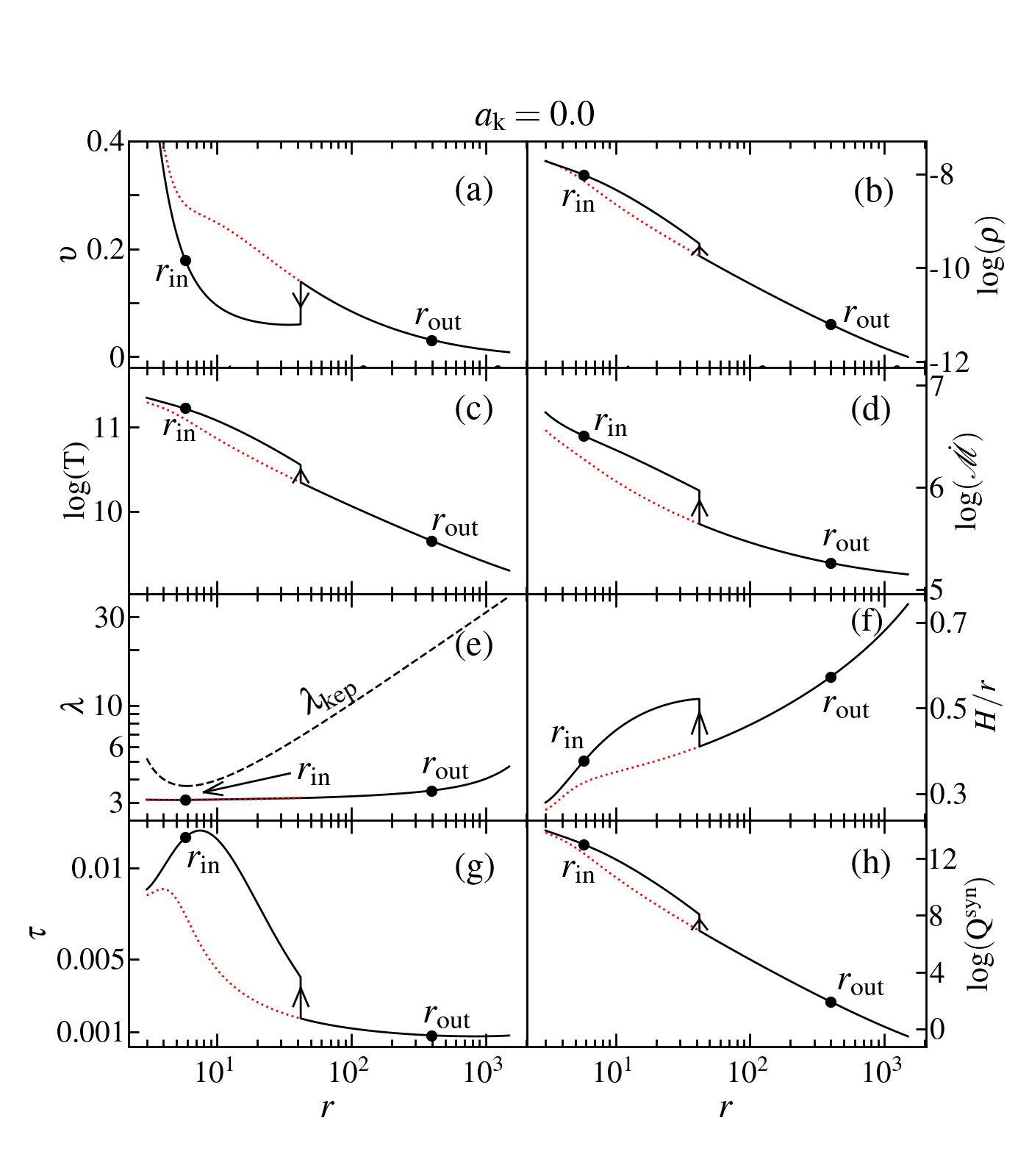}
		\caption{Plot of (a) radial flow velocity $\upsilon$, (b) density $\rho$, (c) Temperature $T$, (d) entropy accretion rate ${\dot {\cal M}}$, (e) angular momentum $\lambda$, (f) disk aspect ratio $H/r$, (g) optical depth $\tau$, and (h) synchrotron emissivity $Q^{\rm syn}$ for shocked accretion solution with $r$. Here, the vertical arrows indicate the shock transition location. Dotted curve refers shock free solution. See the text for details.
		}
		\label{fig:4}
	\end{center}
\end{figure}

In figure \ref{fig:4}, we present the profile of the different flow variables for the shocked accretion solution presented in figure \ref{fig:3}. We depict the variation of the radial velocity ($\upsilon$) in Figure \ref{fig:4}(a), where discontinuous transition of $\upsilon$ is observed at the shock radius ($r_s$). We show the radial variation of mass density ($\rho$) of accreting matter in figure \ref{fig:4}(b) and find that $\rho$ increases monotonically with the decrease of $r$ in the pre-shock branch although sudden jump of $\rho$ is yielded across the shock front. Such a density jump at $r_s$ is inevitable in order to maintain the conservation of mass flux (see equation 2). Because of this, PSC experiences density compression which is eventually quantified in terms of compression ratio defined as $R=\Sigma_{+}/\Sigma_{-}$, where $\Sigma~(=2 \rho H)$ is vertically integrated mass density of accretion flow at a given radial coordinate. We obtain $R=2.31$. In figure \ref{fig:4}(c), the variation of temperature ($T$ in Kelvin) with $r$ is shown. Indeed, the temperature of PSC shoots up as the kinetic energy of the upstream (pre-shock) flow is transformed into thermal energy in the downstream (post-shock) flow, which eventually resulted the increase of PSC temperature. Usually, the temperature jump at $r_s$ is determined by means of the shock strength ($S$), and it is defined as $S=M_{-}/M_{+}$, where $M_{-} ~ (M_{+})$ being the pre-shock (post-shock) Mach number. We obtain $S=2.92$. We present the entropy accretion rate (${\dot {\cal M}}$) in figure \ref{fig:4}(d) and show that ${\dot {\cal M}}$ at PSC is larger compared to pre-shock region. This discernibly indicates that the shock-induced GTAS are favourable over the shoch-free GTAS according to the second law of thermodynamics \cite[]{Becker-Kazanas2001}. We demonstrate angular momentum ($\lambda$) variation in figure \ref{fig:4}(e) and find that the transport of $\lambda$ remains feeble within several hundreds of gravitational radius, although it increases rapidly towards the outer edge of the disk ($r_{\rm edge}$). This possibly happens as viscous time-scale becomes larger than infall time-scale of accretion flow around BH. We show that disc thickness is scaled with radial coordinate in figure \ref{fig:4}(f) and find that $H/r < 1$ is maintained all throughout ($r_h \lesssim r \le r_{\rm edge}$) in presence of shock. Furthermore, we display the variation of scattering optical depth $\tau$ in figure \ref{fig:4}g. In this work, $\tau$ is given by $\tau = \kappa \rho H$, where $\kappa =0.38$ cm$^2$g$^{-1}$. Since  $\tau < 1$ particularly at $r < r_s$, the disc continues to remain as optically thin there. Hence, the hard X-ray radiations originated from PSC would escape significantly with ease. In Fig. \ref{fig:4}h, we present the synchrotron emissivity (in ${\rm erg}~{\rm cm}^{-3}~{\rm s}^{-1}$) with $r$. From the figure, it is evident that the net energy loss from PSC is highly profound in comparison with pre-shock flow.

\begin{figure}
	\begin{center}
		\includegraphics[width=0.9\textwidth]{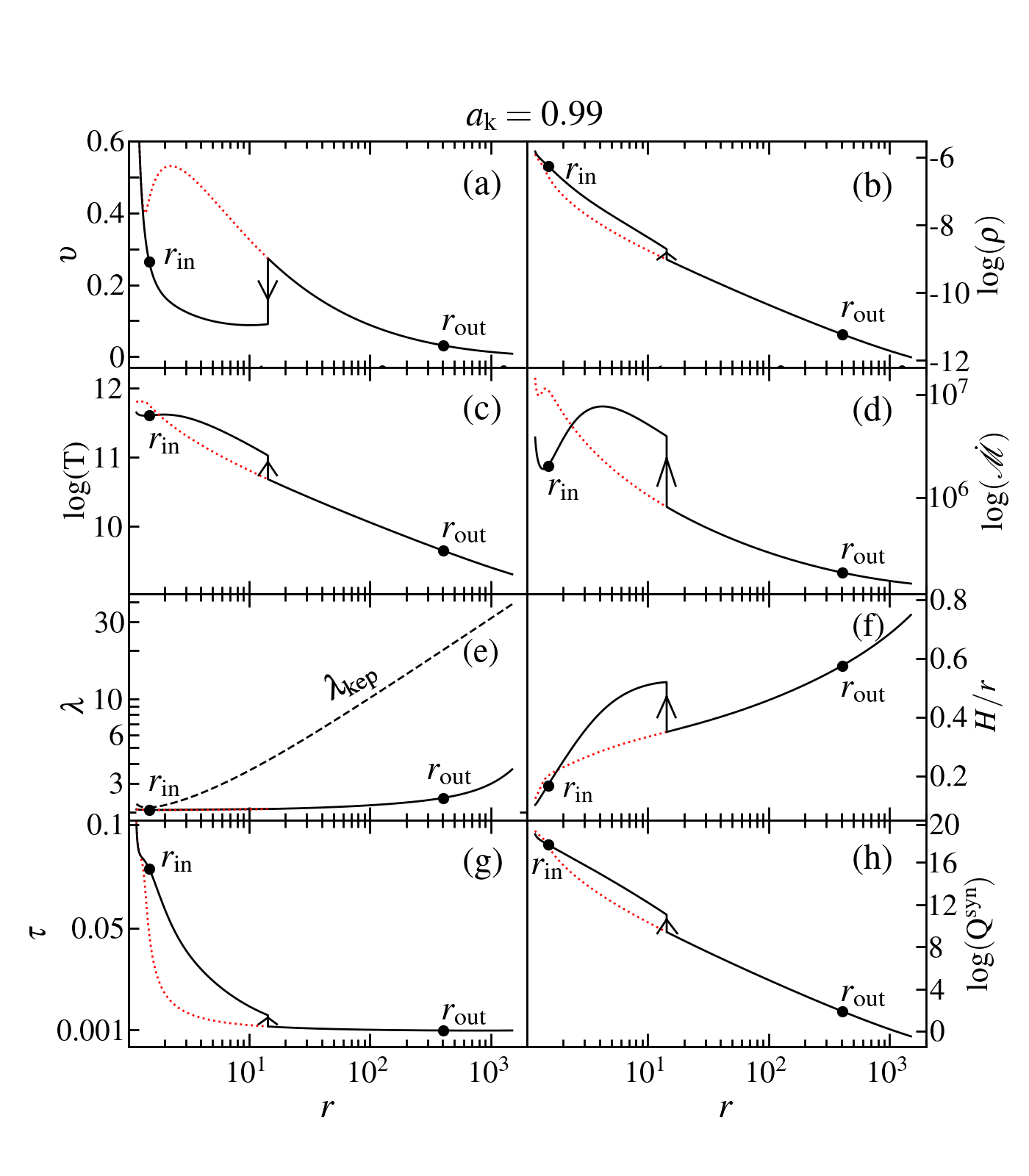}
		\caption{ Same as Fig. \ref{fig:4}, but for flow accreting onto a rotating BH of $a_{\rm k}=0.99$. 
		}
		\label{fig:5}
	\end{center}
\end{figure}

In a similar way, we depict the different flow variables, such as $\upsilon,~\rho,~T,~{\dot {\cal M}},~\lambda, ~H/r, ~\tau$ and $Q^{\rm syn}$ in figure \ref{fig:5} corresponding to the shocked accretion flow around a spinning BH of spin $a_{\rm k}=0.99$ presented in figure \ref{Eq:3}b. Figure evidently indicates that the overall radial variations of these quantities are qualitatively similar with the results of $a_{\rm k}=0.0$, except the region at the vicinity of the BH horizon ($r_h$). In particular, we find that $\tau$ continues to increase as the flow approaches to the horizon ($r \rightarrow r_h$), although it is seen to decrease for weakly rotating BH (see Fig. \ref{fig:4}). This happens because $\tau$ is broadly regulated by the density ($\rho$) and $\rho$ is increased significantly at the vicinity of BH having $a_{\rm k}=0.99$. We skip the detail descriptions of other quantities to avoid repetitions.

\begin{table*}
	\caption{Black hole spin ($a_{\rm k}$), power law exponent ($\theta$), inner critical point location ($r_{\rm in}$), inner critical point angular momentum ($\lambda_{\rm in}$), outer critical point location ($r_{\rm out}$), outer critical point angular momentum ($\lambda_{\rm out}$), shock radius ($r_s$), compression ratio ($R$), and shock strength ($S$) for global transonic solutions presented in figure \ref{fig:6}. See text for more details.
	}
	\begin{center}\label{table-3}
		\begin{tabular}{l l c c c c c c c l   }\hline\hline
			$a_{\rm k}$ & $\theta$ & $r_{\rm in}$ & $\lambda_{\rm in}$ & $r_{\rm out}$  & $\lambda_{\rm out}$ & $r_s$ & $R$ & $S$   \\ 
			& & $(r_g)$    & $(r_g c)$      & $(r_g)$     &$(r_g c)$ & $(r_g)$    & &  \\ \hline			
			0 & 0 & 5.315 & 3.222 & 386.573 & 3.419 & 129.30 & 1.65 & 1.88 \\
			& 0.01 & 5.520 & 3.162 & 387.529 & 3.373 & 69.57 & 2.07 & 2.51 \\
			& 0.02 & 5.787 & 3.098 & 388.552 & 3.322 & 40.71 & 2.35 & 2.99 \\
			& 0.033 & 6.295 & 3.007 & 389.986 & 3.252 & 17.58 & 2.54 & 3.35 \\ 
			\hline			
			0.99 & 0 & 1.529 & 2.122 & 601.228 & 2.368 & 35.78 & 2.90 & 4.10 \\
			& 0.01 & 1.612 & 2.076 & 601.479 & 2.342 & 18.15 & 3.17 & 4.78 \\
			& 0.015 & 1.668 & 2.052 & 601.615 & 2.329 & 13.13 & 3.24 & 4.99 \\
			& 0.021 & 1.828 & 2.001 & 601.907 & 2.300 & 6.67 & 3.27 & 5.11 \\ \hline
				
		\end{tabular}
	\end{center}
\end{table*}

\begin{figure}
	\begin{center}
		\includegraphics[width=0.7\textwidth]{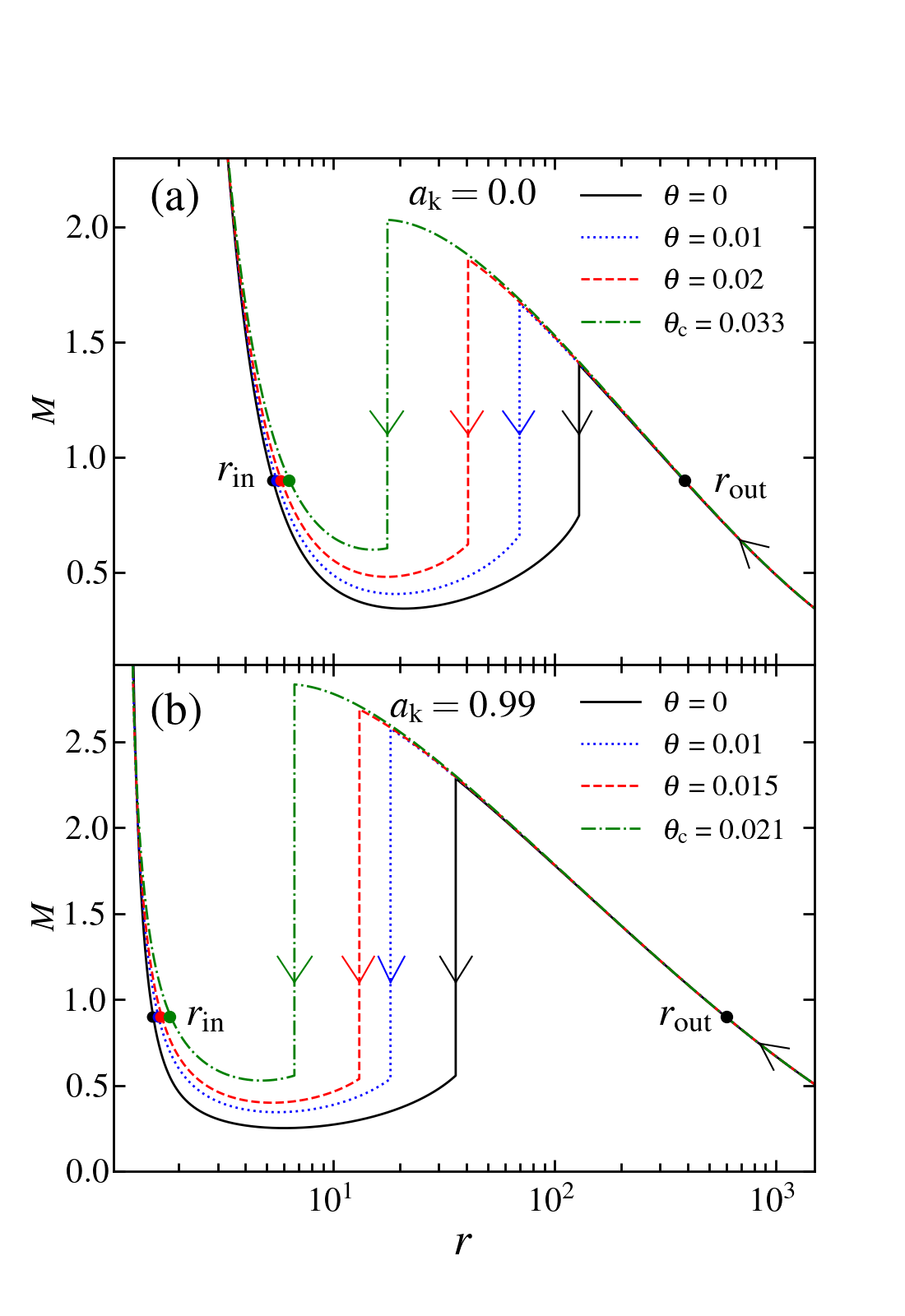}
		\caption{Plot of $M$ with $r$ for accretion solutions that contain shock waves. Here, global parameters are chosen as ${\dot m} = 0.01$, and $\alpha_0=0.01$. {\it Top panel:} Flows with ${\cal E}_{\rm edge} = 1.0004$ and $\lambda_{\rm edge} = 4.01$ are injected from $r_{\rm edge} = 1500$ with different $\theta$ values onto a stationary BH of $a_{\rm k}=0.0$. {\it Bottom panel:} Flows with ${\cal E}_{\rm edge} = 1.00023$ and $\lambda_{\rm edge} = 2.67$ at $r_{\rm edge} = 1500$ are injected with different $\theta$ values onto a rotating BH of $a_{\rm k}=0.99$. Critical points are shown using filled circles and vertical down-arrows represent shock transition radii. See the text for details.}
		\label{fig:6}
	\end{center}
\end{figure}

In figure \ref{fig:6}, we display how shock radius changes with $\theta$ values for flows injected with fixed outer boundary values at $r_{\rm edge}$. Here, we choose ${\dot m} = 0.01$, and $\alpha_0=0.01$. In the top panel, we set the energy ${\cal E}_{\rm edge} = 1.0004$ and angular momentum $\lambda_{\rm edge} = 4.01$ at $r_{\rm edge} = 1500$ and allow the flow to accrete onto a non-rotating BH of $a_{\rm k}=0.0$. We note that for $\theta=0.0$, the subsonic flow becomes supersonic at $r_{\rm out}=386.573$ and shock is formed at $r_s=129.30$ as the RH conditions are satisfied there. We also calculate compression ratio as well as shock strength for this solution and obtain as $R=1.65$ and $S=1.88$. This solution is shown with solid curve, whereas solid vertical arrow denotes shock radius. As $\theta$ is increased to $\theta=0.01$, shock front moves inwards at $r_s = 69.57$. This happens due to the fact that the increase of $\theta$ enhances the viscous effect in the accretion flow, and hence, the transport of $\lambda$ in the outward direction becomes more intense. This effectively weakens the centrifugal repulsion resulting the shock to move closer to the BH horizon. Evidently, this finding suggests that shock formation in accretion flow is centrifugally driven. Here, we obtain $R=2.07$ and $S=2.51$. We plot this solution using dotted curve. For the purpose of representation, we plot another solution for $\theta = 0.02$ using dashed curve. Indeed, the value of $\theta$ can not be increased indefinitely, and we find that beyond a limiting value of $\theta$, which is $\theta_c=0.033$, RHC for shock are not favourable and hence, shock does not form. Interestingly, time-varying shock may still be possible, however, investigation of this is beyond the scope of this paper. Note that $\theta_c$ does not owns a universal value as it is dependent on other flow variables. Accretion solution for $\theta_c=0.033$ are depicted using dot-dashed curve. In the bottom panel, we present the shocked accretion solutions for flows accreting onto rotating BH of $a_{\rm k}=0.99$. Here, we choose energy ${\cal E}_{\rm edge} = 1.00023$ and angular momentum $\lambda_{\rm edge} = 2.67$ at $r_{\rm edge} = 1500$. The solutions depicted with solid, dotted, dashed and dot-dashed correspond to $\theta = 0$, $0.01$, $0.015$, and $\theta_c=0.021$, respectively. In Table \ref{table-3}, we tabulate the flow quantities corresponding to these accretion solutions harbouring shock waves. 

\begin{figure}
	\begin{center}
		\includegraphics[width=0.7\textwidth]{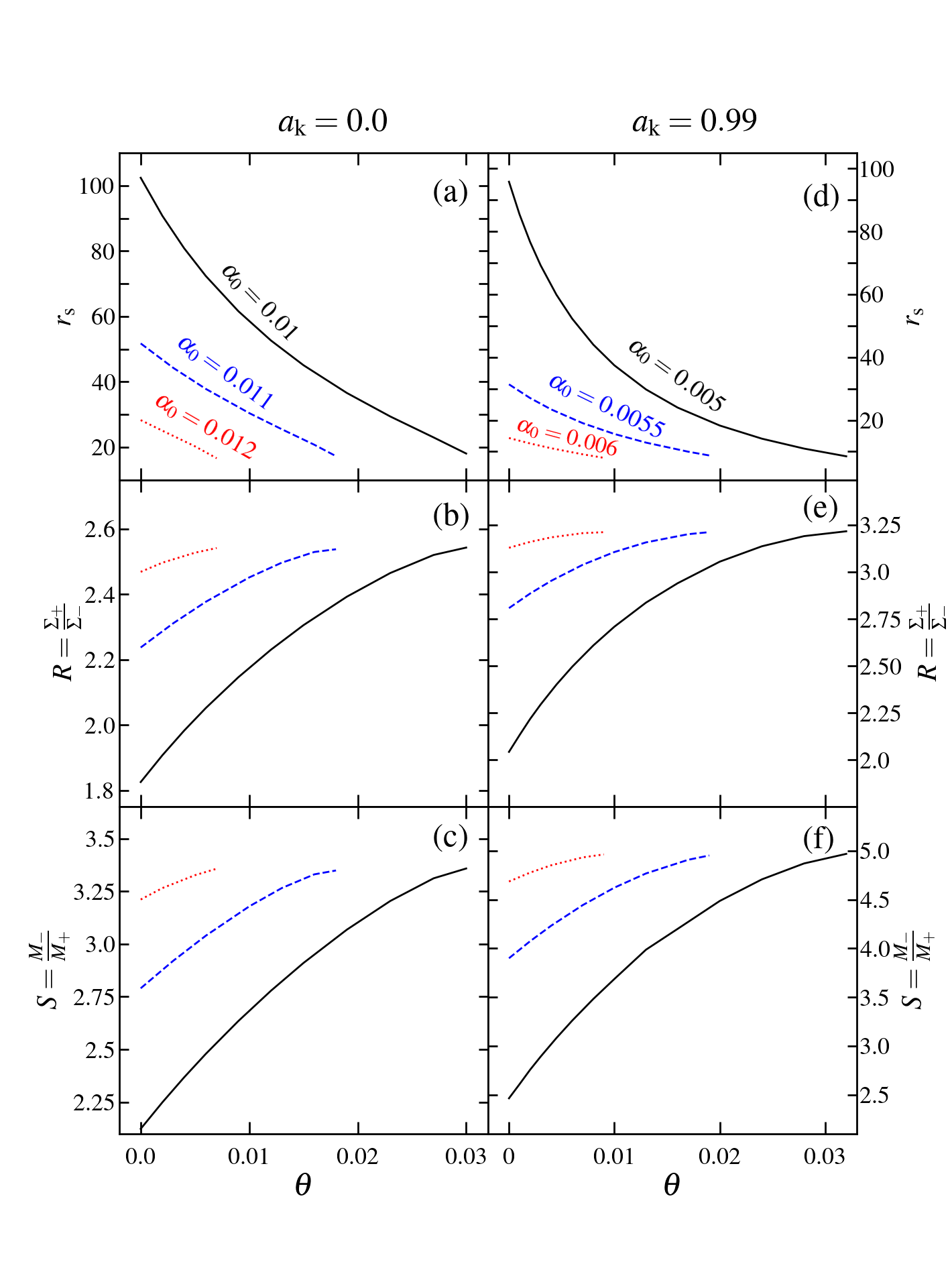}
		\caption{Comparison of shock location $r_s$ (upper panel), compression ratio $R$ (middle panel), and shock strength $S$ (lower panel) when varied with $\theta$. Here, flows with same energy and angular momentum are injected from the fixed outer edge ($r_{\rm edge}$). {\it Left panels:} Results are obtained for $a_{\rm k}=0$, and solid, dashed, and dotted curves are drawn for $\alpha_0 = 0.01$, $0.011$ and $0.012$, respectively. {\it Right panels:} Results are for $a_{\rm k}=0.99$, where solid, dashed, and dotted curves correspond to $\alpha_0 = 0.005$, $0.0055$ and $0.006$, respectively. See the text for details.}
		\label{fig:7}
	\end{center}
\end{figure}

In figure \ref{fig:7}, variation of shock properties, namely shock radius $r_s$ (upper panel), compression ratio $R$ (middle panel), and shock strength $S$ (lower) are depicted with $\theta$. In the left panels, we display the results for the stationary BH of $a_{\rm k}=0.0$, where flows are injected from $r_{\rm edge}=1500$ with identical energy (${\cal E}_{\rm edge}=1.0004$) and angular momentum ($\lambda_{\rm edge}=3.98$). Here, we set ${\dot m}=0.01$ and obtain the results for $\alpha_0 = 0.01$ (solid) , $0.011$ (dashed) and $0.012$ (dotted), respectively. Figure \ref{fig:7}a clearly shows that stable shocks exist for an ample range of $\theta$ values. As already anticipated, for a fixed $\alpha_0$, $r_s$ decreases with $\theta$ as it weakens the centrifugal repulsion against the gravitational attraction. Moreover, for a given $\theta$, when $\alpha_0$ is higher, the angular momentum transport becomes more efficient weakening the centrifugal barrier. Because of this, shock front proceeds inwards. Notice that for a fixed $\alpha_0$, when $\theta > \theta_c$, shock disappears as RH conditions are not satisfied. As indicated earlier that $\theta_c$ strictly depends on the other flow variables (see \S4). Indeed, the radiative cooling processes that primarily determines the flux of the high energy radiations from the disk are strongly dependent on both $\rho$ and $T$ distributions across shock front \cite[]{Chakrabarti-Titarchuk1995,Mandal-Chakrabarti2005}. Keeping this in mind, in figure \ref{fig:7}b, we depict the variation of the compression ratio ($R$, measure of density compression across shock) as function of $\theta$ corresponding to shock-induced GTAS presented in figure \ref{fig:7}a. We observe that when $\theta$ is increased, shock is generally pushed towards the BH. Due to this, PSC becomes further compressed causing the overall increase of $R$. Similar trend is generally observed in the variation of $R$ irrespective of the $\alpha_0$ values provided shock exists. Similarly, in figure \ref{fig:7}c, we display how shock strength ($S$, measure of temperature jump across the shock front) varies with $\theta$ for the solutions presented in figure \ref{fig:7}a. It is clear that for a fixed $\alpha_0$, shock strength $S$ monotonically increases as $\theta$ is increased and ultimately shifted from weaker to stronger regime. We continue the analyses and present the outcome for $a_{\rm k}=0.99$ in the right side panels of figure \ref{fig:7}, where flows are injected from $r_{\rm edge}=1500$ with identical ${\cal E}_{\rm edge}=1.0004$, $\lambda_{\rm edge}=2.44$, and ${\dot m}=0.001$. In order to preserve $\theta$ range intact, here we choose relatively smaller ${\dot m}$ compared to the same used for flows around weakly spinning BH. In figures \ref{fig:7}d-f, results are are obtained for $\alpha_0 = 0.005$ (solid), $0.0055$ (dashed) and $0.006$ (dotted), respectively. Note that the overall variations of $r_s$, $R$, and $S$ with $\theta$ for $a_{\rm k}=0.99$ appear qualitatively similar as delineated in the left panels for $a_{\rm k}=0.0$.

\begin{figure}
	\begin{center}
		\includegraphics[width=0.7\textwidth]{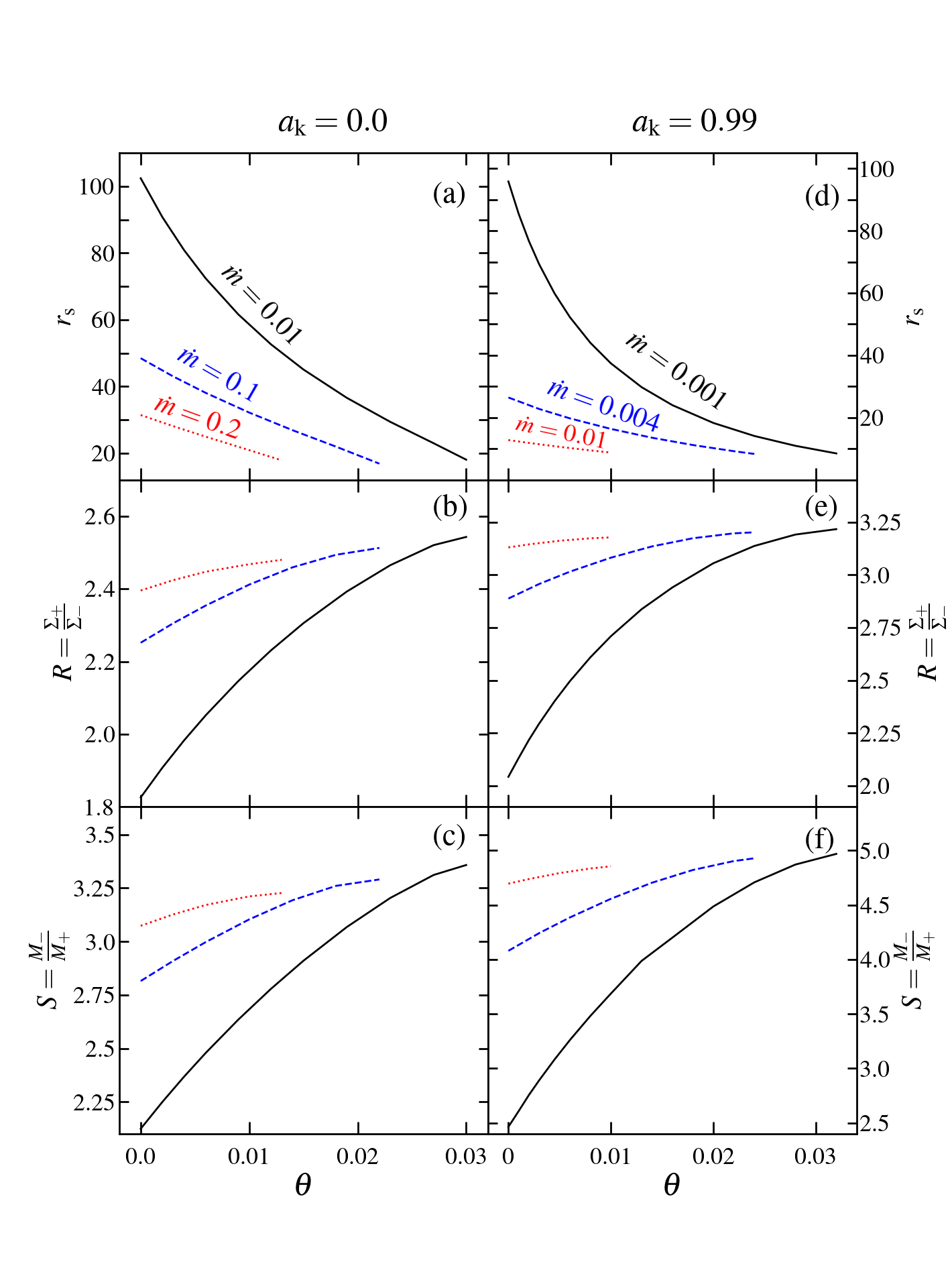}
		\caption{Plot of $r_s$ (upper panel), $R$ (middle panel), and $S$ (lower panel) with $\theta$ for flows with different ${\dot m}$. {\it Left panels:} Here, we choose $r_{\rm edge}=1500$, ${\cal E}_{\rm edge}=1.0004$, $\lambda_{\rm edge}=3.98$, and $a_{\rm k}=0.0$. Solid, dashed, and dotted curves are for ${\dot m} = 0.01$, $0.1$ and $0.2$. {\it Right panels:} We set $r_{\rm edge}=1500$, ${\cal E}_{\rm edge}=1.0004$, $\lambda_{\rm edge}=2.44$, and $a_{\rm k}=0.99$. Solid, dashed, and dotted curves are for ${\dot m} = 0.001$, $0.004$ and $0.01$. See the text for details.}
		\label{fig:8}
	\end{center}
\end{figure}

In figure \ref{fig:8}, we investigate the effect of accretion rate (${\dot m}$) for shock triggering in a convergent accretion flow. Such a portentous effort is very much useful as the radiative cooling processes are regulated by ${\dot m}$. While doing so, we inject matter onto a non-rotating BH ($a_{\rm k}=0.0$) from $r_{\rm edge}=1500$ with ${\cal E}_{\rm edge}=1.0004$, $\lambda_{\rm edge}=3.98$ and $\alpha_0=0.01$. We display the obtained results in left panels, where solid, dashed, and dotted curves correspond to ${\dot m} = 0.01$, $0.1$ and $0.2$, respectively. Similarly, for $a_{\rm k}=0.99$, we choose $r_{\rm edge}=1500$, ${\cal E}_{\rm edge}=1.0004$, $\lambda_{\rm edge}=2.44$, $\alpha_0=0.005$, and results are drawn in the right panels. The spin values are marked at the top of the figure. In panels (a) and (d), we present the variation of $r_s$ with $\theta$, where shocks are seen to proceed further inward close to BH horizon as $\theta$ increases. This feature is commonly observed irrespective to ${\dot m}$ values provided shock is formed. What is more is that for a given $\theta$, when ${\dot m}$ is increased, shock front moves inward. This is not surprising because higher ${\dot m}$ eventually increases the effect of cooling in PSC, and accordingly, thermal pressure decreases. Hence, shock settles down at the location closer to the horizon to maintain pressure balance on both sides of the discontinuity. In panels (b) and (e), we compare the compression ratio ($R$) and notice that $R$ increases with $\theta$. This evidently indicates that for a convergent flow, accretion shock becomes stronger as $r_s$ decreases. We further find that shock strength $S$ increases monotonically with $\theta$, and for a given $\theta$, when shocks form closer to BH, $S$ is enhanced and vice versa (see panels (c) and (f)).

\begin{figure}
	\begin{center}
		\includegraphics[width=0.8\textwidth]{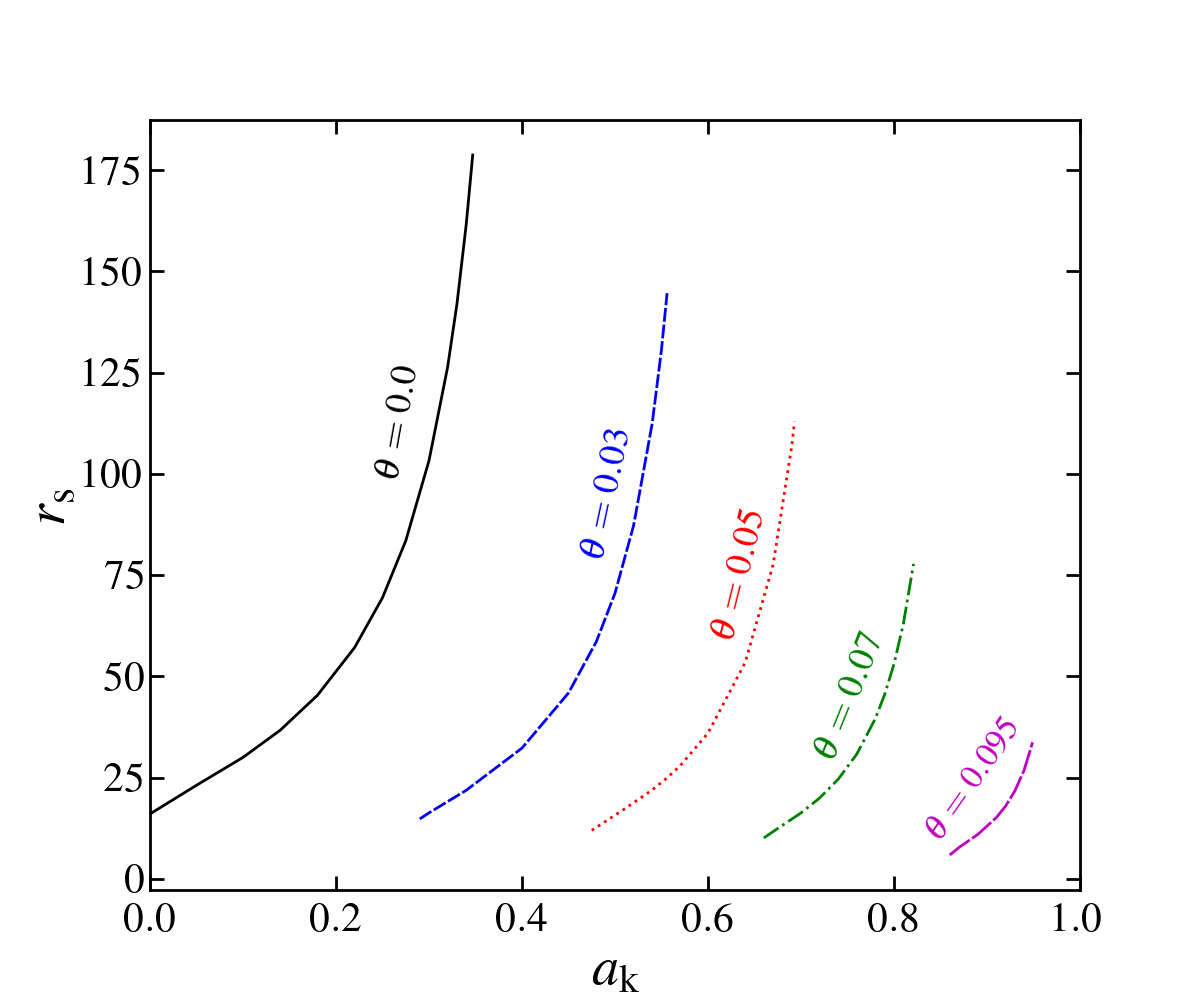}
		\caption{Variation of $r_{\textrm{s}}$ with $a_{\textrm{k}}$ for different $\theta$ values. Results plotted using solid, dashed, dotted, dot-dashed and big-dashed curves are for $\theta = 0.0, 0.03, 0.05, 0.07$, and $0.095$. See the text for details.}
		\label{fig:9}
	\end{center}
\end{figure}

Next, we investigate the effect of BH rotation ($a_{\rm k}$) on $r_s$ and present the obtained results in figure \ref{fig:9}, where the variation of $r_s$ with $a_{\rm k}$ for different $\theta$ is depicted. For this analysis, we inject matter from $r_{\rm edge} = 1500$ with $\lambda_{\rm edge} = 3.78$, ${\cal E}_{\rm edge} = 1.0004$, ${\dot m} = 0.01$ and $\alpha_{0} = 0.01$. In the figure, solid, dashed, dotted, dot-dashed and big-dashed curves are used to indicate results correspond to $\theta=0.0, 0.03, 0.05, 0.07$, and $0.095$, respectively. It is clear from the figure that for a fixed $\theta$, $r_s$ moves outwards from BH horizon as $\theta$ increases for flows with fixed outer boundary conditions. Accordingly, the effective size of the PSC is increased, and hence, the possibility of up-scattering the soft-photons from pre-shock disk at PSC is increased in producing the high energy radiations. We further notice that for a given $\theta$, shocks form for a particular range of $a_{\rm k}$, and as $\theta$ is increased, the range of $a_{\rm k}$ is shifted to the higher side. This is not surprising because of the fact that for a fixed $\lambda_{\rm edge}$, higher $\theta$ increases angular momentum transport causing the overall reduction of $\lambda (r)$ close to BH. Indeed, it is evident that for higher $a_{\rm k}$, shock exists when $\lambda$ is relatively low \cite[]{Das-Chakrabarti2008}, and this happens because of spin-orbit coupling present in the effective potential (see equation (1a)) describing spacetime geometry around BH. These findings are consistent with the results of \cite{Sen-etal2022}. In contrary, we observe that $r_s$ decreases due to the increase of $\theta$ for flows accreting onto a BH having a fixed spin ($a_{\rm k}$) value. Moreover, we observe that the lower limit of $r_s$ is gradually reduced when the flow with fixed outer boundary accretes onto the BHs of increasing spin ($a_{\rm k}$) values.

\section{Shock parameter space}

\begin{figure}
	\begin{center}
	\includegraphics[width=0.49\textwidth]{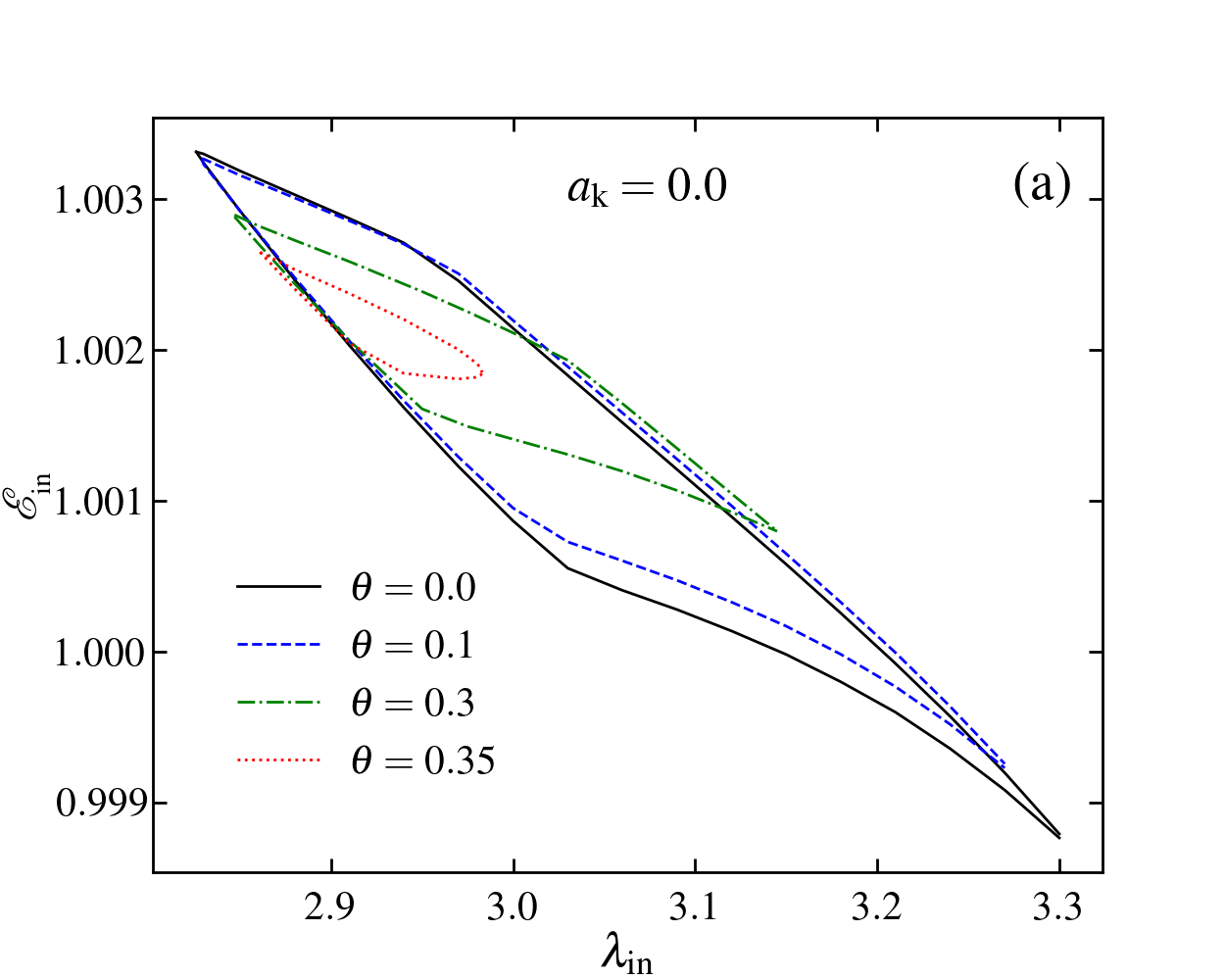}
	\includegraphics[width=0.49\textwidth]{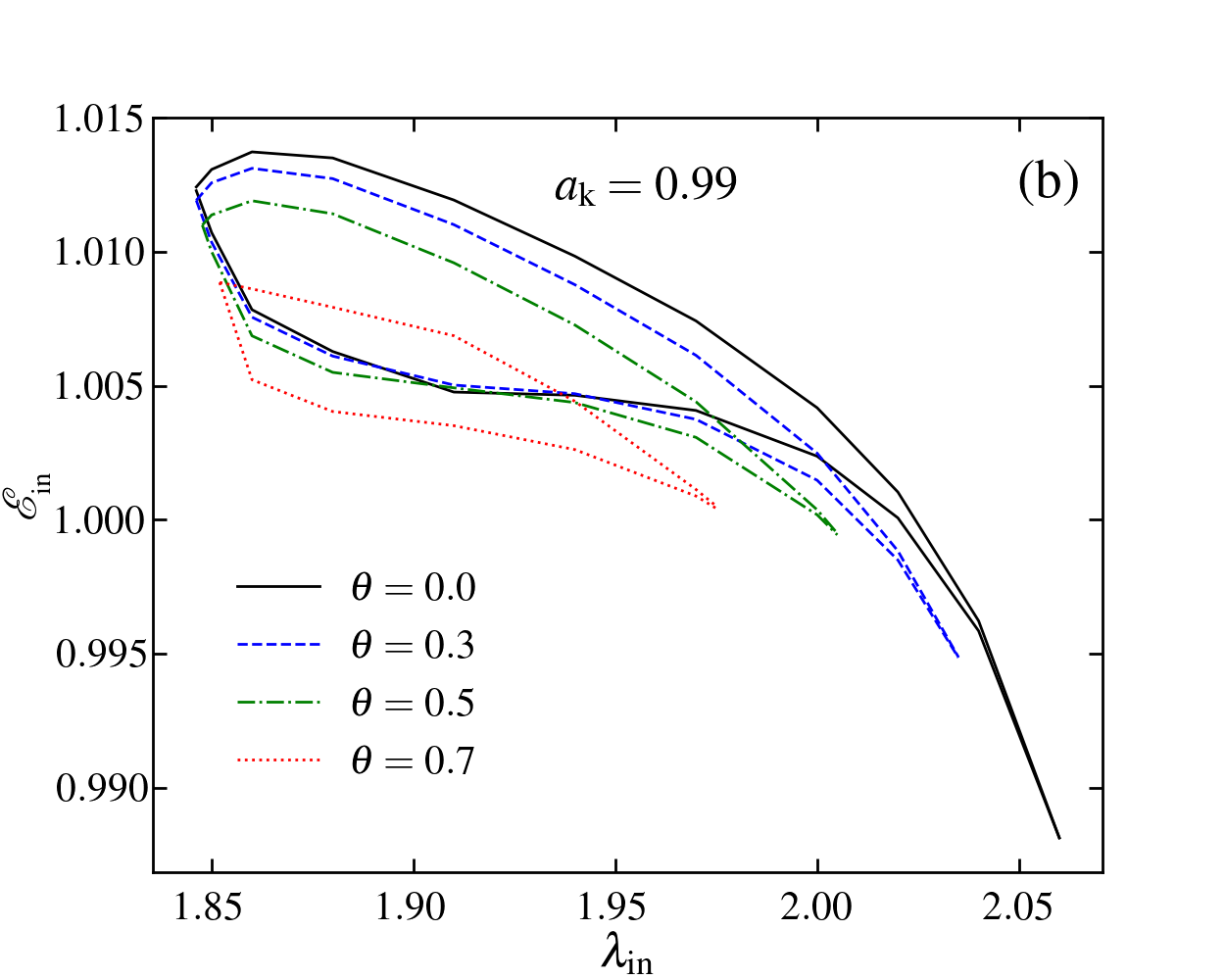} 
	\caption{Effective domain of the shock parameter space as function of $\theta$. Upper panel (a) is for the stationary BH of $a_{\rm k} = 0$, whereas bottom panel (b) illustrates results for the rapidly rotating BH of $a_{\textrm{k}} = 0.99$. See the text for details.}
	\label{fig:10}
	\end{center}
\end{figure}

In this section, we proceed further to identify the region of parameter space that admits stationary shock solutions for viscous advective accretion flow around BHs. It is evident from figures \ref{fig:6}-\ref{fig:9} that shocked-induced GTAS are obtained for a range of angular momentum and $\theta$ values. Hence, we examine how the shock properties alters with $\theta$ in a viscous flow, and classify the effective domain of parameter space in terms of $\theta$ in the $\lambda_{\rm in}-{\cal E}_{\rm in}$ plane, where $\lambda_{\rm in}$ and ${\cal E}_{\rm in}$ refer to the angular momentum and energy of the flow at $r_{\rm in}$ \cite[]{Aktar-etal2017,Aktar-etal2019}. We choose ${\cal E}_{\rm in}$ and $\lambda_{\rm in}$ in defining the shock parameter space as the flow is expected to advect into BH with energy and angular momentum resembling these values. The results are presented in figure \ref{fig:10}, where left panel is for $a_{\rm k}=0.0$ and the effective region bounded with solid, dashed, dot-dashed, and dotted curves are for $\theta = 0.0, 0.1, 0.3$, and $0.35$, respectively. Here, we set ${\dot m}=0.01$. Similarly, in the bottom panel, we illustrate the results for $a_{\rm k}=0.99$, where solid, dashed, dot-dashed, and dotted curves are used to separate the region for $\theta = 0.0, 0.3, 0.5$, and $0.7$, respectively. Here, we choose ${\dot m}=10^{-4}$. In each panel, $a_{\rm k}$ and $\theta$ values are marked. We observe that in both panels, the effective domain of $\lambda_{\rm in}-{\cal E}_{\rm in}$ space for standing shock is reduced as $\theta$ increases. and accordingly, the shock formation possibility is also diminished \cite[]{Chakrabarti-Das2004,Das-2007}. Indeed, when $\theta$ exceeds its limiting value ($i.e.$, $\theta > \theta^{\rm max}$), the parameter space for standing shock disappears. Note that for $\theta > \theta^{\rm max}$, flow angular momentum at the vicinity of BH is reduced to such a limit that the centrifugal barrier becomes very weak and it could not trigger the shock transition. Hence, standing shock ceases to exist. Nevertheless, time-dependent shocked accretion solutions may exist for $\theta > \theta^{\rm max}$, which were examined by numerical simulation to study the oscillatory behaviour of shock solutions \cite[]{Molteni-etal1994,Das-etal2014,Sukova-Janiuk2015,Lee-etal2016}. Interestingly, the solutions of this kind satisfactorily account for the quasi-periodic oscillations (QPOs) phenomenon that are commonly observed in BH-XRBs \cite[]{Nandi-etal2012,Sreehari-etal2020,Majumder-etal2022}. However, we indicate that the study of time-dependent shock solution is beyond the scope of this framework and we plan to consider it as future work.

\begin{figure}
	\begin{center}
		\includegraphics[width=0.9\textwidth]{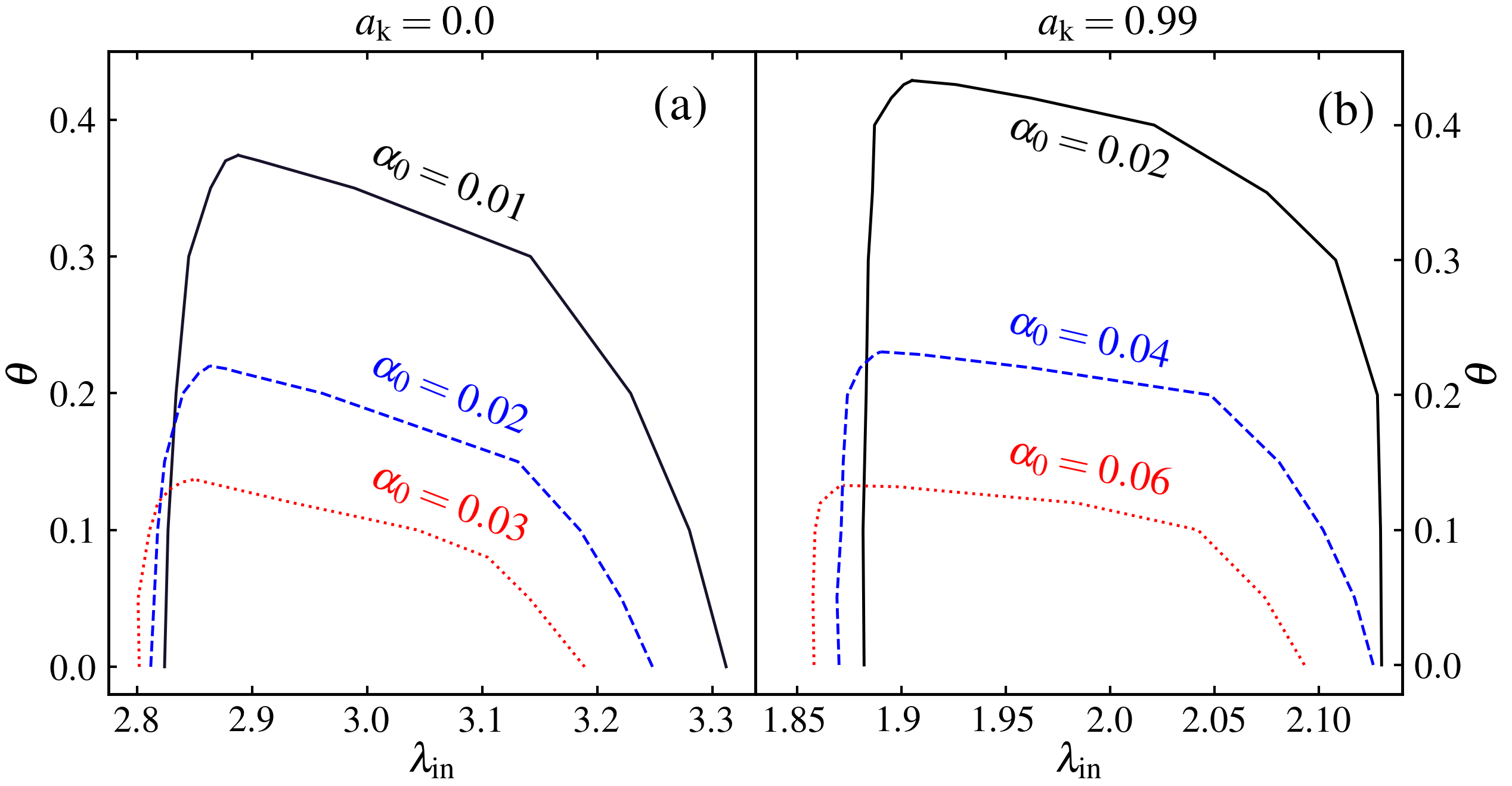}
		\caption{Variation of $\theta$ with angular momentum at the inner critical point ($\lambda_{\rm in}$) that admits shock. In panel (a), $a_{\rm k} = 0.0$ and solid, dashed and dotted curves denote results for $\alpha_0 = 0.01, 0.02$, and $0.03$, respectively. Similarly, in panel (b), $a_{\rm k} = 0.99$, and solid, dashed and dotted curves are for $\alpha_0 = 0.02, 0.04$, and $0.06$. Here, we choose $\dot m=0.01$. See the text for details.}
		\label{fig:11}
	\end{center}
\end{figure}

We continue our study to examine the ranges of $\lambda_{\rm in}$ and $\theta$ in terms of $\alpha_0$ that admit shocked-induced GTAS. In order to do that, we set ${\dot m}=0.01$ and scan the range of $\theta$ for a given set of ($\lambda_{\rm in},\alpha_0$) by freely varying $r_{\rm in}$ (equivalently ${\cal E}_{\rm in}$). The obtained results for $a_{\rm k}=0.0$ and $0.99$ are shown in figure \ref{fig:11}. In panel (a), solid, dashed, and dotted are obtained for $\alpha_0=0.01$, $0.02$, and $0.03$, that separate the region for shocked accretion solutions from the shock-free solutions. Similarly, in panel (b), solid, dashed, and dotted are obtained for $\alpha_0=0.02$, $0.04$, and $0.06$. We observe that the permissible region for shock in $\lambda_{\rm in}-\theta$ plane gradually diminishes with the increase of $\alpha_0$ for both slowly and rapidly rotating BHs. In addition, we find that for a given $\alpha_0$, $\theta$ attains its maximum value, namely $\theta^{\rm max}$, at a fixed $\lambda_{\rm in}$.  We further observe that as $\alpha_0$ is increased, the value of $\theta^{\rm max}$ is decreased and it is obtained at smaller $\lambda_{\rm in}$ values.

\begin{figure}
	\begin{center}
		\includegraphics[width=0.7\textwidth]{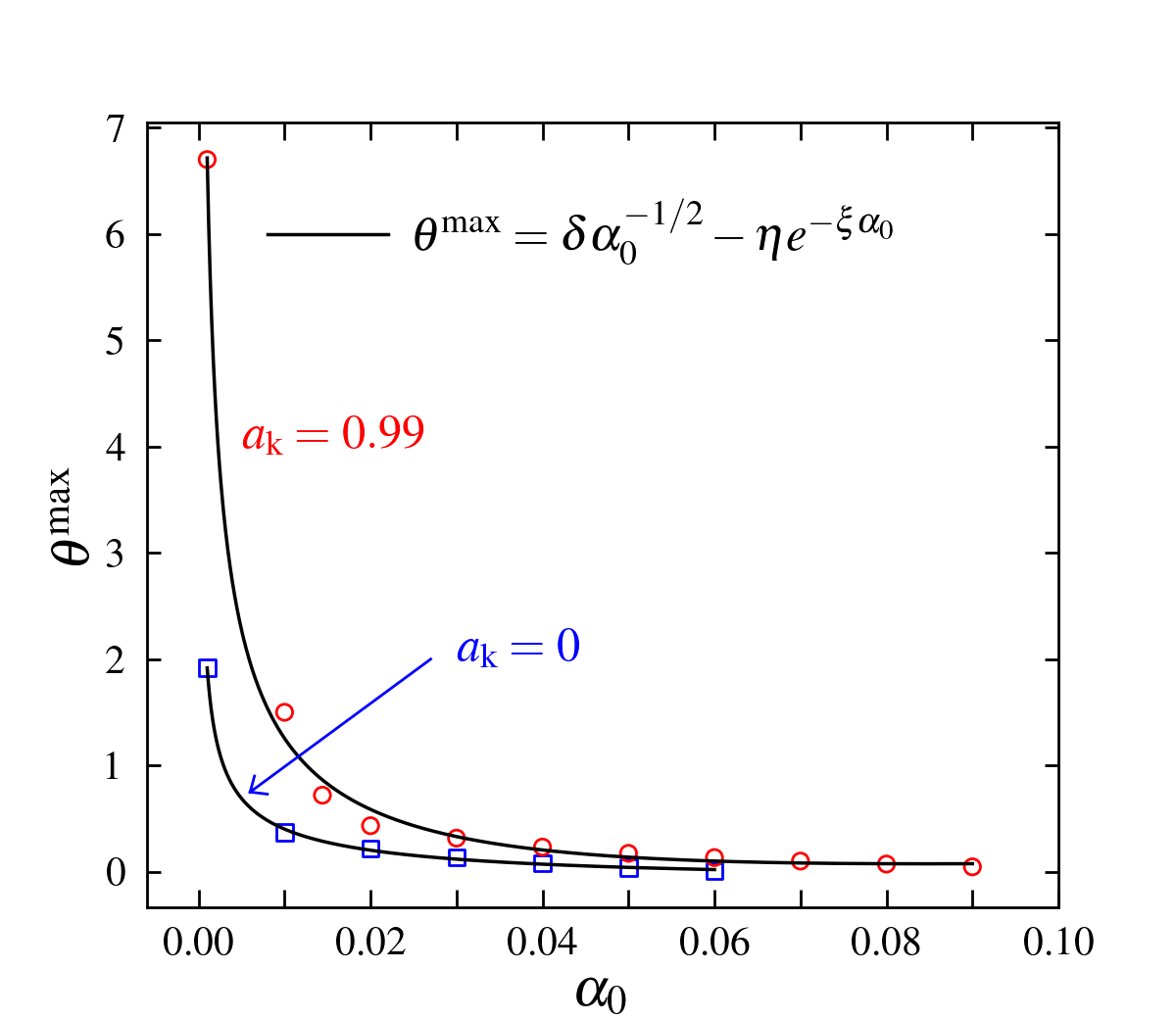}
		\caption{Plot of $\theta^{\rm max}$ as function of $\alpha_{0}$ for shocked accretion solutions. Results depicted using open squares and open circles are for $a_{\rm k}=0.0$ and $0.99$, respectively. Solid curves denote the fitted function as mentioned in the text. Here, we choose ${\dot m}=0.01$. See the text for details.
		}
		\label{fig:12}
	\end{center}
	
\end{figure}

\begin{table*}
	\caption{Values of the coefficients obtained from the best fit representation of $\theta^{\rm max} ~(= \delta \alpha^{-1/2}_{0} - \eta e^{-\xi \alpha_0})$ yielding shock-induced GTAS (see figure \ref{fig:12}). See the text for details.}
	\centering
	\begin{tabular}{c c c c } 
		\hline \hline
		$a_{\rm k}$ & $\delta$ & $\eta$  & $\xi$   \\ \hline
		$0$         & $0.07$   & $0.31$ & $2.50$ \\
		$0.99$      & $0.25$   & $1.39$ & $6.48$ \\
		\hline \hline
	\end{tabular}
	\label{table-4}
\end{table*}

In figure \ref{fig:12}, we demonstrate how $\theta^{\rm max}$ varies with $\alpha_0$. Open squares represent the results for $a_{\rm k}=0.0$, while open circles are for $a_{\rm k}=0.99$. These data points are further fitted empirically as $\theta^{\rm max}= \delta \alpha^{-1/2}_{0} - \eta e^{-\xi \alpha_0}$, where $\delta$, $\eta$, and $\xi$ are the constants, and their values strictly depends on $a_{\rm k}$ which are presented in Table \ref{table-4}. In the figure, solid curves denote the best-fit representations of the fitted function described above for $a_{\rm k}=0.0$ and $0.99$, respectively. Figure clearly indicates that the accretion flows with relatively higher viscosity continue to harbour shock waves around the highly spinning BHs as compared to the weakly rotating BHs. This happens mostly because of the fact that the outer critical points turn into nodal type \cite[]{Dihingia-etal2019a} at higher viscosity for $a_{\rm k} \rightarrow 0$, and hence, shock-induced GTAS ceases to exist.

\section{Conclusions}

In this study, we examine the structure of the viscous accretion flow that includes the more general viscosity than those usually discussed in the literature \cite[and references therein]{Narayan-Yi1994,Chakrabarti1996,Chakrabarti-Das2004}. In particular, we consider the viscosity parameter to vary with the radial coordinate as $\alpha(r)$ and observe that GTAS continue to exist around rotating BH. Depending on input parameters, $i.e.$, viscosity, angular momentum, accretion rate, the accretion flow may harbour shock waves. Indeed, the shock-induced GTAS is promising in the sense that it has the potential to explain the spectro-temporal properties of BH sources \cite[]{Chakrabarti-Manickam2000,Nandi-etal2001,Smith-etal2001,Smith-etal2002,Nandi-etal2012,Iyer-etal2015,Das-etal2021}. We find important results that are presented below.

\begin{enumerate}
	\item There exists global transonic accretion solutions either pass through inner critical points ($r_{\rm in}$, ADAF-type) or outer critical points ($r_{\rm out}$, Bondi type) for low angular momentum flow. We find that when viscosity is appropriately chosen by tuning the viscosity exponent $\theta$, keeping the other flow parameter fixed at the outer edge ($r_{\rm edge}$), ADAF-type solutions change its character to become Bondi type (see figure \ref{fig:1}). Further, when $\theta$ is decreased for an ADAF-type solution, global accretion solution eventually becomes closed as it could not extend upto the disk outer edge (see figure \ref{fig:2}), although it can join with a Bondi-type solution via Rankine-Hugoniot shock transition (see figure \ref{fig:3}). Note that these findings are seen in both weakly rotating ($a_{\rm k}\rightarrow 0$) and rapidly rotating ($a_{\rm k}\rightarrow 1$) BHs. Since it is generally perceived that BHs may accrete low angular momentum matter from its surroundings stars, shock seems to be an indispensable component in the accretion flow. 
	
	\item We observe that because of shock transition, convergent accretion flow is compressed yielding hot and dense PSC (see figure \ref{fig:4}-\ref{fig:5}). Thus, PSC contains swarm of hot electrons, which are likely to reprocess the low energy photons from pre-shock flow via inverse-Comptonization and {generate hard X-ray radiations} \cite[]{Chakrabarti-Titarchuk1995,Mandal-Chakrabarti2005}. Such a signature of excess high energy radiations is often observed from galactic X-ray binaries harbouring BH sources \cite[and references therein]{Sunyaev-Titarchuk1980,Iyer-etal2015,Baby-etal2020}. With this, we infer that the shock radius ($r_s$) which coarsely measures the size of PSC seems to play viable role to emit hard X-ray radiations from accretion disc. 
	
	\item When viscosity is enhanced, the efficiency of the angular momentum transport increases that evidently weakens the centrifugal repulsion against gravity. Because of this, for higher $\theta$, the size of PSC is reduced as the shock front  moves inward to satisfy the pressure equilibrium on both sides of the discontinuity (see figure \ref{fig:6}).  Accordingly, by suitably changing $\theta$, one can regulate the accreting dynamics including PSC while explaining the disk emission.
			
	\item We determine the limiting range of flow parameters	that admit shock transition in viscous accretion flow around both slowly and rapidly rotating BHs. We find that shock-induced GTAS are not discrete solutions. In fact, solutions of this kind are obtained for ample range of the flow parameters (see figure \ref{fig:10}). However, the possibility of shock formation diminishes as we increase viscosity, and beyond a critical limit of $\theta > \theta^{\rm max}$, shock disappears. Indeed, $\theta^{\rm max}$ does not owns a universal value as it is dependent on other flow variables.
		
	\item We quantify $\theta^{\rm max}$ as function of $\alpha_0$ for $a_{\rm k}=0$ and $0.99$, and find that it sharply decreases at lower $\alpha_0$ and ultimately settles down to its asymptotic limit (see figure \ref{fig:12}).
	
	\end{enumerate}

It is noteworthy to refer that in the literature, there exists results of shock-induced transonic accretion flows obtained from simulation studies \cite[]{Chakrabarti-Molteni1995,Lanzafame-etal1998,Giri-Chakrabarti2012,Das-etal2014,Okuda-Das2015,Lee-etal2016}. Indeed, in all these works, the viscosity parameter ($\alpha$) was treated as global constant all throughout the disk. On the contrary, adopting the variable viscosity prescription, numerical simulation results of accretion flows around BHs are also reported. In particular, \cite{Hawley-Krolik2001,Hawley-Krolik2002} examined the dynamical behaviour of the azimuthally and time averaged $\alpha$ that typically ranges between $\sim 0.01$ and $\sim 0.1$ throughout most of the disk. \cite{Penna-etal2013} reported feeble variation of $\alpha$ ($\sim 0.01-0.3$) across the disk length scale with a peak around $2-3r_g$. In studying truncated accretion disk, \cite{Hogg-Reynolds2018} obtained $\alpha \sim 0.07$ at $\sim 600r_g$ in the quasi-steady state, although $\alpha$ settles down to $\approx 0.02$ at the inner edge of the disk. Evidently, in these variable $\alpha$ studies, the formation of shock is not observed simply because these simulations were performed with Keplerian or quasi-Keplerian flows which are subsonic in nature and hence, are incapable of triggering shock transition \cite[]{Das-etal2001a}. Accordingly, it remains infeasible to compare the results obtained from the present formalism with the existing simulations. Nevertheless, we infer that with the suitable choices of the input parameters, accretion flow having variable $\alpha$ would possibly be capable in possessing shock as corroborated in \cite{Giri-Chakrabarti2013}. We further indicate that based on the the above findings, the quantitative description of the viscosity profile adopted in the present formalism seems to be fairly consistent with the results of the simulation works. 

We further mention that in an accreting system, PSC seems to play vital role in deciphering the observational signatures commonly observed in BH X-ray binary sources. As indicated earlier that PSC can reprocesses the soft photons via inverse Comptonization to produce hard X-ray radiations which eventually contributes in generating the high energy tail of the energy spectrum \cite[]{Chakrabarti-Titarchuk1995,Mandal-Chakrabarti2005}. Occasionally, Galactic X-ray binaries do show spectral state transitions, which is possibly resulted when PSC geometry alters \cite[and references therein]{Nandi-etal2012, Iyer-etal2015}. When PSC demonstrates time varying modulation, it resembles an astonishing phenomenon known as Quasi-periodic Oscillations (QPO) of hard X-ray radiations \cite[]{Chakrabarti-Manickam2000,Nandi-etal2001,Das-etal2014}. Moreover, it has been reported that PSC can deflect a part of the accreting matter in the form of jets/outflow \cite[]{Das-etal2001b,Das-etal2014,Aktar-etal2017,Aktar-etal2018,Nandi-etal2018}. Considering all these, we argue that the present formalism in examining the PSC characteristics is fervently relevant in the astrophysical context.

Finally, we indicate the limitations of this formalism as it is developed considering several assumptions. We use effective potential to describe the space-time geometry around the rotating BH avoiding rigorous general relativistic approach. We neglect structured large scale magnetic fields and use stochastic magnetic field
configuration. We also consider the flow to remain confined in single temperature domain although flow is expected to maintain two-temperature (for both ions and electrons) profiles. Implementation of all such issues is beyond the scope of this work and we intend to take up these relevant issues in future projects.

\section*{Data Availability}

The data underlying this article will be available with reasonable request.

\section*{Acknowledgements}

This work was supported by the Science and Engineering Research Board (SERB) of India through grant MTR/2020/000331.



\begin{appendices}

\section{Detail expression of the Wind equation}

With some simple algebraic steps, the radial momentum equations, azimuthal momentum equations and entropy generation equations are reduced to the following form as,
\begin{equation}
R_{0}+ R_{\upsilon}\frac{d \upsilon}{d r} + R_{\Theta}\frac{d \Theta}{d r} + R_{\lambda}\frac{d \lambda}{d r} = 0,
\end{equation}

\begin{equation}
L_{0} + L_{\upsilon}\frac{d \upsilon}{d r} + L_{\Theta}\frac{d \Theta}{d r} + L_{\lambda}\frac{d \lambda}{d r} = 0 ,
\end{equation}

\begin{equation}
E_{0} + E_{\upsilon}\frac{d \upsilon}{d r} + E_{\Theta}\frac{d \Theta}{d r} + E_{\lambda}\frac{d \lambda}{d r} = 0.
\end{equation}

Using the equations (A1-A3), we obtain the wind equation, derivative of angular momentum and derivative of temperature which are given by,
\begin{equation}
\frac{d\upsilon}{d r} = \frac{\cal N}{\cal D},
\end{equation}
\begin{equation}
\frac{d\lambda}{d r} = \lambda_{1}+\lambda_{2}\frac{d\upsilon}{d r},
\end{equation}
\begin{equation}
\frac{d\Theta}{d r} = \Theta_{1}+\Theta_{2}\frac{d\upsilon}{d r}, 
\end{equation}
where,
\begin{equation*}
\begin{aligned}
{\cal N} ={} & E_{\lambda}\left(-R_{\Theta} L_{0}+ R_{0} L_{\Theta}\right) + E_{\Theta}\left(R_{\lambda} L_{0}- R_{0} L_{\lambda}\right)  \\
& + E_{0}\left(-R_{\lambda} L_{\Theta}+ R_{\Theta} L_{\lambda}\right) ,
\end{aligned}
\end{equation*}

\begin{equation*}
\begin{aligned}
{\cal D} ={} &  E_{\lambda}\left(R_{\Theta} L_{\upsilon}+ R_{\upsilon} L_{\Theta}\right) + E_{\Theta} \left(-R_{\lambda} L_{\upsilon}+ R_{\upsilon} L_{\lambda}\right)  \\
& + E_{\upsilon}\left(R_{\lambda} L_{\Theta} -  R_{\Theta} L_{\lambda}\right),
\end{aligned}
\end{equation*}

\begin{equation*}
\begin{aligned}
\Theta_{1} = & \frac{\Theta_{11}}{\Theta_{33}} \thickspace \thickspace \Theta_{2} = \frac{\Theta_{22}}{\Theta_{33}} ,\thickspace\thickspace \lambda_{1} = \frac{\lambda_{11}}{\Theta_{33}}, \thickspace\thickspace \lambda_{2} = \frac{\lambda_{22}}{\Theta_{33}},
\end{aligned}
\end{equation*}

\begin{equation*}
\begin{aligned}
\Theta_{11} = &	E_{\lambda} L_{0}- E_{0} L_{\lambda},\thickspace\thickspace \Theta_{22} = E_{\lambda} L_{\upsilon} - E_{\upsilon} L_{\lambda},\thickspace \Theta_{33} = - E_{\lambda} L_{\Theta} + E_{\Theta} L_{\lambda} ,
\end{aligned}
\end{equation*}

\begin{equation*}
\begin{aligned}
\lambda_{11} = 	-E_{\Theta} L_{0}+ E_{0} L_{\Theta},\thickspace\thickspace \lambda_{22} = -E_{\Theta} L_{\upsilon} + E_{\upsilon} L_{\Theta},
\end{aligned}
\end{equation*}

\begin{equation*}
\begin{aligned}
R_{0} =& \frac{d \Phi_{e}^{\rm{eff}}}{d r} - \frac{3\Theta}{r \tau h} + \frac{F_{1}\Theta}{\tau {\cal F} h} - \frac{\Theta \Delta^\prime}{\tau \Delta h},\thickspace R_{\Theta} = \frac{1}{\tau h}, \tau = 1 + \frac{m_p}{m_e},\\
R_{\lambda} = & \frac{F_{2} \Theta}{\tau {\cal F} h },\thickspace R_{\upsilon} = \upsilon -\frac{2\Theta}{\tau \upsilon h}, \thickspace \Delta^\prime = \frac{d\Delta}{dr},
\end{aligned}
\end{equation*}

\begin{equation*}
\begin{aligned}
E_{0} = &-\frac{Q^{-}}{\rho} -r \alpha \upsilon^{2}\omega_{1} - \frac{2 r \alpha\Theta \omega_{1}}{\tau} + \frac{\upsilon \Theta\left(-r F_{1}\Delta + {\cal F}(3\Delta + r \Delta') \right)}{r\tau\Delta},
\end{aligned}
\end{equation*}

\begin{equation*}
\begin{aligned}
E_{\Theta} = & \frac{\left(1+ 2 n \upsilon\right)}{\tau},\thickspace E_{\lambda} - \frac{\left(F_{2} \upsilon \Theta +  r \alpha {\cal F}( \tau \upsilon^{2} + 2\Theta)\omega_{2}\right)}{\tau {\cal F}},\thickspace E_{\upsilon} = \frac{2\theta}{\tau},
\end{aligned}
\end{equation*}

\begin{equation*}
\begin{aligned}
L_{0}= & -2\alpha \upsilon^{2} - \frac{4\alpha\Theta}{\tau} + \frac{r \alpha \upsilon^{2} \Delta'}{2 \Delta} + \frac{r \alpha\Theta\Delta^\prime}{\tau \Delta} - \frac{r}{\tau}\left(\tau\upsilon^{2}+2\Theta\right)\frac{d\alpha}{dr},
\end{aligned}
\end{equation*}

\begin{equation*}
\begin{aligned}
L_{\Theta} = & -\frac{2 r \alpha}{\tau},\thickspace L_{\lambda} = \upsilon,\thickspace L_{\upsilon}  = -r \alpha \upsilon + \frac{2 r \alpha \Theta}{\tau \upsilon},
\end{aligned}
\end{equation*}

\begin{equation*}
\begin{aligned}
F_{1} = & \frac{F \lambda \omega_{1}}{(1-\lambda\Omega)^{2}} +  \frac{1}{1-\lambda\Omega}\frac{dF}{dr},
\end{aligned}
\end{equation*}

\begin{equation*}
\begin{aligned}
F_{2} = \frac{F \Omega}{(1-\lambda\Omega)^{2}} + \frac{F \lambda \omega_{2}}{(1-\lambda\Omega)^{2}},\thickspace {\cal F} = \frac{1}{(1-\lambda\Omega)} F ,
\end{aligned}
\end{equation*}

\begin{equation*}
\begin{aligned}
F = \frac{(r^{2}+a_{\textrm{k}}^{2})^{2} + 2 \Delta a_{\textrm{k}}^2}{(r^{2}+a_{\textrm{k}}^{2})^{2} - 2 \Delta a_{\textrm{k}}^2}, \thickspace \frac{d {\cal F}}{dr} = F_{1} + F_{2}\frac{d \lambda}{d r},\thickspace \frac{d \Omega}{d r} = \omega_{1}+\omega_{2}\frac{d \lambda}{d r},
\end{aligned}
\end{equation*}

\begin{equation*}
\begin{aligned}
\omega_{1} = -\frac{2\left(a_{\textrm{k}}^{3} + 3 a_{\textrm{k}} r^{2}+ \lambda(a_{\textrm{k}} \lambda - 2 a_{\textrm{k}}^{2} + r^{2}(r-3))\right)}{\left(r^{3} + a_{\textrm{k}}^2 (r+2) - 2 a_{\textrm{k}} \lambda\right)^{2}},
\end{aligned}
\end{equation*}

\begin{equation*}
\begin{aligned}
\omega_{2} = \frac{r^{2}\left(a_{\textrm{k}}^2 + r (r - 2)\right)}{\left(r^{3} + a_{\textrm{k}}^2 (r+2) - 2 a_{\textrm{k}} \lambda\right)^{2}}.
\end{aligned}
\end{equation*}

\end{appendices}


\end{document}